\documentclass[sigconf, nonacm]{acmart}
\setcopyright{none}
\newcommand\vldbavailabilityurl{}
\newcommand\vldbpagestyle{plain}

\usepackage{amsthm}
\usepackage{graphicx}
\usepackage{subcaption}
\usepackage{xspace}
\usepackage{listings}
\usepackage{fancyvrb}
\usepackage{cleveref}
\usepackage{hyperref}
\hypersetup{
    colorlinks=true,
    linkcolor=blue,
    urlcolor=blue,
    citecolor=blue
}
\usepackage{comment}

\usepackage{markdown}
\usepackage{mathtools} % for \Coloneqq
\usepackage{float}
\newfloat{listing}{htbp}{lol}
\floatname{listing}{Example}
\usepackage[textwidth=1.7cm]{todonotes}
\usepackage[dvipsnames]{xcolor} 
\usepackage{float} % for [H] float option
\usepackage{booktabs}
\usepackage{multirow}
\usepackage{tikz}
\usetikzlibrary{positioning, shapes.geometric}
\usepackage{lineno}
\usepackage{pgfplots}
\pgfplotsset{compat=1.18}
\definecolor{codeblue}{RGB}{0,102,153}      % #006699 for functions
\definecolor{codered}{RGB}{217,74,122}      % #d94a7a for keywords
\definecolor{codepurple}{RGB}{125,71,147}   % #7d4793 for strings
\definecolor{codegray}{RGB}{102,102,102}    % #666666 for comments
\definecolor{codegreen}{RGB}{102,153,0}     % #669900 for special keywords
\definecolor{backcolour}{RGB}{248,248,248}  % Light gray background

\definecolor{RelaNNkw}{RGB}{175,0,219}       % keywords: def, enddef (purple)
\definecolor{RelaNNtype}{RGB}{38,127,153}    % ER/relation names (teal)
\definecolor{RelaNNfunc}{RGB}{121,94,38}     % transform refs, built-in ops (brown)
\definecolor{RelaNNnum}{RGB}{9,134,88}       % numbers (green)
\definecolor{RelaNNtmpl}{RGB}{25,25,112}     % template brackets & params (midnight blue)
\definecolor{RelaNNcontentattr}{RGB}{204,102,0}  % content attributes: s, t (dark orange)
\definecolor{RelaNNembattr}{RGB}{180,30,30}       % embedding attributes: z, w, z_q, ... (dark red)
\definecolor{RelaNNbound}{RGB}{180,30,140}        % bounding constructs: Join, Set, Union, | (magenta)

\newcommand{\dslfontsize}{\small}
\newcommand{\dslwrap}[1]{\ifmmode#1\else{\dslfontsize\ttfamily#1}\fi}
\newcommand{\kw}[1]{\dslwrap{\textcolor{RelaNNkw}{\textbf{#1}}}}   % def, enddef
\newcommand{\rn}[1]{\dslwrap{\textcolor{RelaNNtype}{#1}}}          % relation / ER names
\newcommand{\fn}[1]{\dslwrap{\textcolor{RelaNNfunc}{#1}}}          % transforms, built-in ops
\newcommand{\ca}[1]{\dslwrap{\textcolor{RelaNNcontentattr}{#1}}}   % content attributes
\newcommand{\ea}[1]{\dslwrap{\textcolor{RelaNNembattr}{#1}}}       % embedding attributes
\newcommand{\tp}[1]{\dslwrap{\textcolor{RelaNNtmpl}{#1}}}          % template brackets & params
\newcommand{\nm}[1]{\dslwrap{\textcolor{RelaNNnum}{#1}}}           % numbers, hyper-params
\newcommand{\bd}[1]{\dslwrap{\textcolor{RelaNNbound}{#1}}}         % bounding constructs
\newcommand{\cmt}[1]{\dslwrap{\textcolor{codegray}{#1}}}            % comments
\newcommand{\neck}{\dslwrap{\textcolor{RelaNNkw}{\textbf{:-}}}}    % rule neck separator
\newcommand{\nrcode}[1]{\texttt{#1}}                            % code identifiers in prose (DataFrame, cuDF, ...)
\newcommand{\dslinline}[1]{\dslwrap{#1}}                        % inline DSL fragments in prose: matches the code listings (\small typewriter; use semantic color macros inside)
\newcommand{\repjoin}{``$\mathtt{,\dots}$''}                    % join replicator in prose
\newcommand{\repunion}{``$\mathtt{|\dots}$''}                   % union replicator in prose

\DefineVerbatimEnvironment{code}{Verbatim}{%
  commandchars=\\\{\},
  fontsize=\small,
  xleftmargin=8pt,
  xrightmargin=4pt,
  formatcom={\normalfont\ttfamily\def\dslfontsize{}},
  baselinestretch=0.95
}

\DefineVerbatimEnvironment{codesmall}{Verbatim}{%
  commandchars=\\\{\},
  fontsize=\small,
  xleftmargin=6pt,
  xrightmargin=2pt,
  formatcom={\normalfont\ttfamily\def\dslfontsize{}},
  baselinestretch=0.95
}
\usepackage{etoolbox}
\BeforeBeginEnvironment{codesmall}{\vspace{-4pt}}
\AfterEndEnvironment{codesmall}{\vspace{-4pt}}

\lstdefinestyle{RelaNNstyle}{
    basicstyle=\ttfamily\small,
    breaklines=true,
    keepspaces=true,
    showspaces=false,
    showstringspaces=false,
    tabsize=2,
    xleftmargin=4pt,
    xrightmargin=4pt,
    keywordstyle=\color{codeblue},
    commentstyle=\color{codegray},
    stringstyle=\color{codepurple},
    morecomment=[l]{//},
    morecomment=[s]{/*}{*/},
    morestring=[b]"
}

\lstdefinestyle{RelaNNfigstyle}{
    basicstyle=\ttfamily\small,
    breaklines=true,
    keepspaces=true,
    showspaces=false,
    showstringspaces=false,
    tabsize=4,
    xleftmargin=4pt,
    xrightmargin=4pt,
    commentstyle=\color{codegray},
    morecomment=[l]{//},
}

\newcommand{\ctag}[2]{\tikz[baseline=(c.base)]\node[circle,fill=#1,inner sep=0.5pt,minimum size=1.8ex,font=\sffamily\fontsize{6}{6}\selectfont,text=white](c){#2};}
\newcommand{\tagA}{\ctag{RelaNNtype}{1}}
\newcommand{\tagB}{\ctag{RelaNNcontentattr}{2}}
\newcommand{\tagC}{\ctag{RelaNNkw}{3}}
\newcommand{\tagD}{\ctag{RelaNNfunc}{4}}
\newcommand{\tagE}{\ctag{RelaNNtype}{5}}
\newcommand{\tagF}{\ctag{RelaNNcontentattr}{6}}
\newcommand{\tagG}{\ctag{RelaNNkw}{8}}
\newcommand{\taga}{\ctag{RelaNNtype}{a}}
\newcommand{\tagb}{\ctag{RelaNNcontentattr}{b}}
\newcommand{\tagc}{\ctag{RelaNNkw}{c}}
\newcommand{\tagd}{\ctag{RelaNNfunc}{d}}
\newcommand{\tagH}{\ctag{RelaNNfunc}{7}}

\newcommand{\Linear}{\mathsf{Linear}}
\newcommand{\Concat}{\mathsf{Concat}}

\theoremstyle{definition}

\newcommand{\inlinenote}[3]{\todo[color=#3!60,inline]{\textbf{#2:} #1}}

\newcommand{\mgmargindone}[1]{}

\newcommand{\mginlinedone}[1]{}

\newcommand{\ylinline}[1]{\inlinenote{#1}{YL}{orange}}

\newcommand{\df}{:=}
\newcommand{\join}{\bowtie}
\newcommand{\concat}{\oplus}

\usepackage[normalem]{ulem}

\tikzstyle{process} = [rectangle, minimum width=2cm, minimum height=1cm, text centered, draw=black, fill=orange!20]
\def\vl#1{\texttt{#1}}
\def\emb#1{\tikz\draw[black,fill=#1] (0,0) circle (.5ex);}

\def\reals{\mathbb{R}}
\def\e#1{\emph{#1}}
\def\angs#1{\mathord{\left\langle#1\right\rangle}}

\def\embpart#1{{\color{MidnightBlue} #1}}
\def\embangs#1{\embpart{\angs{#1}}}

\def\scs{\mathcal{S}}

\newcommand*{\ldblbrace}{\{\mskip-5mu\{}
\newcommand*{\rdblbrace}{\}\mskip-5mu\}}

\newcommand{\lmset}{\ldblbrace}
\newcommand{\rmset}{\rdblbrace}

\def\bag#1{\mathord{\ldblbrace#1\rdblbrace}}

\def\rel#1{\mbox{\dslfontsize\ttfamily\textcolor{RelaNNtype}{#1}}}  % relation names in prose/figures/math: render like the DSL listings (small typewriter, teal). \mbox keeps it valid inside math expressions; relation names carry no subscripts.

\newcommand{\eat}[1]{}

\def\e#1{\emph{#1}}

\newcommand{\rneck}{\mathrel{\,{:}\mbox{--}\,}}

\long\def\yl#1{{\color{blue}YL: #1}}

\title{Incorporating Deep Learning Design in Database Queries}
\author{Yuval Lev Lubarsky}
\authornote{Equal contribution.}
\affiliation{%
  \institution{Technion}
  \city{Haifa}
  \country{Israel}}
\email{lubarsky@cs.technion.ac.il}

\author{Dean Light}
\authornotemark[1]
\affiliation{%
  \institution{University of Washington}
  \city{Seattle}
  \country{USA}}
\email{deanlcs@cs.washington.edu}

\author{Boaz Berger}
\affiliation{%
  \institution{Technion}
  \city{Haifa}
  \country{Israel}}
\email{boazb@campus.technion.ac.il}

\author{Shunit Agmon}
\affiliation{%
  \institution{Technion}
  \city{Haifa}
  \country{Israel}}
\email{shunita@campus.technion.ac.il}

\author{Benny Kimelfeld}
\affiliation{%
  \institution{Technion \& RelationalAI}
  \city{Haifa}
  \country{Israel}}
\email{bennyk@cs.technion.ac.il}

\begin{document}
\begin{abstract}
Deep learning over relational databases is conventionally realized by translating data into graph representations and applying graph-based neural networks within external frameworks %\yl{, before mapping the learned representations back into the database}
.
%before mapping learned representations back into the database. 
This round-trip between the database and external machine learning (ML) systems introduces non-trivial engineering overhead. %data must be extracted, reformatted, fed through separate toolchains, and reintegrated. 
In effect, these graph neural networks operate on tuple embeddings and manipulate them in ways that capture the interactions induced by relational joins. Given this natural correspondence, there is no fundamental reason why specifying a neural network over relational data should be substantially harder than %writing a query over it.
querying it.
We propose an approach that naturally integrates deep learning with 
%the declarative abstraction of 
database queries. The key idea is to associate each tuple with provenance, represented as a vector embedding
%a vector embedding 
%defined by symbolic expressions over 
with learnable parameters. Queries are lifted to operate jointly on data and embeddings, mapping input relations with embedded tuples to output relations with embedded tuples. This approach provides a declarative foundation for relational deep learning, facilitating integration with database systems, optimization, and wide adoption. 
%We establish a precise formalization of this framework through an extension of the relational model and a Datalog-style query language. 

We describe RelaNN, a proof-of-concept implementation of this approach built on top of PyTorch and cuDF.
We illustrate the utility of
%the expressiveness and practicality of 
RelaNN by implementing various graph-learning models, including graph convolutional networks, heterogeneous graph transformers,
%. We also implement models that go beyond the expressivity of graph neural networks, including 
hypergraph neural networks and deep homomorphism networks.
% \yl{The latter two require subgraph-pattern enumeration and multi-way interactions that mainstream GNN frameworks such as PyTorch Geometric do not natively support.} 
%RelaNN exhibits performance on the same order of magnitude as established systems on these methods. %showing the practicality of our approach. 
The simplicity of the programs and their competitive runtime performance demonstrate a concrete path toward making the implementation of state-of-the-art neural networks over databases as simple as writing a query.

\end{abstract}

\maketitle

% \yl{VLDB workshop block, verbatim from the official TaDA 2026 template. Prints the required ``VLDB Workshop Reference Format'' line and the CC-BY-NC-ND license / VLDB Endowment footnote.}
% \yl{VLDB reference-format block commented out to save space on page 1 for the May 18 submission. RESTORE for the June 29 camera-ready -- the proceedings pipeline requires it.}
\pagestyle{\vldbpagestyle}
% \begingroup\small\noindent\raggedright\textbf{VLDB Workshop Reference Format:}\\
% \vldbauthors. \vldbtitle. VLDB \vldbyear\ Workshop: \vldbworkshop.\\
% \endgroup
% \begingroup
% \renewcommand\thefootnote{}\footnote{\noindent
%This work is licensed under the Creative Commons BY-NC-ND 4.0 International License. Visit \url{https://creativecommons.org/licenses/by-nc-nd/4.0/} to view a copy of this license. For any use beyond those covered by this license, obtain permission by emailing \href{mailto:info@vldb.org}{info@vldb.org}. Copyright is held by the owner/author(s). Publication rights licensed to the VLDB Endowment. \\
% \raggedright Proceedings of the VLDB Endowment. ISSN 2150-8097. \\
% }\addtocounter{footnote}{-1}\endgroup

\ifdefempty{\vldbavailabilityurl}{}{
\vspace{.3cm}
\begingroup\small\noindent\raggedright\textbf{VLDB Workshop Artifact Availability:}\\
The source code, data, and/or other artifacts have been made available at \url{\vldbavailabilityurl}.
\endgroup
}

\section{Introduction}

Integrating databases and machine learning (ML) is a central challenge in data science. Both paradigms are fundamental to data-driven workflows: databases provide powerful abstractions for modeling, querying, and managing data, while ML enables the extraction of predictive models from it. However, they traditionally rely on different modeling primitives and implementation strategies, making their integration nontrivial. As a result, bridging the gap between databases and ML has been a major challenge in database foundations over the past decades~\cite{DBLP:journals/dagstuhl-manifestos/AbiteboulABBCD018,Papotti2025PanelON}, and has motivated dedicated workshops at leading database conferences. Previous efforts to bridge this gap range from adding black-box ML functionalities to the database~\cite{10.14778/2367502.2367510,kumar2015glm}, through integrating ML building blocks into DBMS kernels~\cite{db4ml2020,li2024gaussml}, to creating unified languages for both operation types~\cite{li2017mlog,schule2019mlearn,Schle2019ML2SQLC,kunft2019lara,Jamil2024TowardAD}.
Advances in deep neural networks, particularly neural networks over graphs~\cite{schlichtkrull2018modeling,wu2020comprehensive,DBLP:journals/tmlr/Muller00R24}, have fundamentally advanced relational deep learning by enabling prediction models to operate directly over structured data, avoiding ad-hoc feature engineering.

Building on graph learning infrastructure, the common approach to applying deep learning over relational databases is to translate the data into a graph representation: records become nodes and foreign-key relationships become edges. A model (e.g., a GNN) is designed and trained over the resulting graph outside of the database context~\cite{DBLP:conf/icml/FeyHHLR0YYL24,DBLP:conf/sigmod/CappuzzoPT20,DBLP:conf/icde/TonshoffFGK23,DBLP:conf/cikm/LubarskyTGK23,chen2025relgnn,cucumides2025augraph}
%\sa{\cite{velickovic2018graph}?} \yl{No, GAT is not a "using gnns to learn over relations", it is rather just a "regular GNN paper"}
. To capture schema information of the database, architectures like GNNs and graph transformers have been extended to \e{heterogeneous} settings, where nodes and edges are associated with different types~\cite{DBLP:conf/aaai/ZhaoWSHSY21,DBLP:conf/www/WangJSWYCY19}. Hence, designing a neural network over a relational database amounts to developing a predictive model over the induced graph using imperative ML graph libraries such as PyTorch Geometric~\cite{fey2019fast}, DGL~\cite{wang2019dgl}, or Scikit-network~\cite{JMLR:v21:20-412}, %\yl{. These libraries are themselves abstractions over PyTorch, providing the graph operations that are cumbersome to write in raw tensor code.}
which act as abstractions over PyTorch, providing the graph operations that are cumbersome to write in raw tensor code. The role of the database query language is reduced to exporting the database into a tensor representation of a graph, and reintegrating the predictions back into the database for further processing.

We argue that this %\sout{software complexity}\yl{round-trip} 
round-trip is unnecessary, and that it arises from an artificial separation of abstractions that forces developers to work across %\sout{separate} \yl{disparate} 
disparate paradigms. Instead, relational query languages are inherently well-suited for designing neural networks over relational databases, given an appropriate extension of the relational model. %\sa{can this be made clearer with an example?} 
From this perspective, applying graph-learning architectures to a database amounts to manipulating \e{tuple embeddings}, where each tuple is associated with a (parametric) numeric vector, and the resulting computations reflect interactions between tuples through relational operations. In particular, similarly to the abstraction of database provenance as semirings~\cite{DBLP:conf/pods/GreenKT07,G21}, 
joins should determine not only how tuples are combined, but also how their embeddings are composed, while projections (grouping) should determine how embeddings are aggregated. By extending the relational model with these embedding semantics, we can express deep-learning architectures directly over the database, without resorting to intermediate graph representations. %This enables neural models to be designed easily using familiar query-language constructs (e.g., SQL and Datalog) in an elegant and declarative manner. 
This allows developers to seamlessly design neural models using familiar, declarative query-language constructs such as SQL and Datalog.
%\dl{I would stress here the reduction of complexity for the person building these intergrated workflows that you talked about in the 1st paragraph}\ylinline{make sure we dont repeat ourselves and its same concept not 2 . maybe cut after break to 2 lines. dont do sharply. do potential of}

As an example, the following program represents query-key-value attention as used by heterogeneous graph transformers~\cite{hu2020heterogeneous}, over a dataset of patients and treatments.
\begin{code}
\rn{Score}(\ca{p},\ca{t}; \ea{q}*\ea{k}) \neck \rn{Treat}(\ca{p},\ca{t}), \rn{Queries}(\ca{p}; \ea{q}), \rn{Keys}(\ca{t}; \ea{k}) .
\rn{Attention}(\ca{p}; \fn{sum}(\ea{a}*\ea{v})) \neck \rn{Score}(\ca{p},\ca{t}; \ea{a}), \rn{Values}(\ca{t}; \ea{v}) .
\end{code}
Each relation is written as $R(\vec{x}; z)$, pairing ordinary content attributes $\vec{x}$ with a tuple \e{embedding} $z$. The first rule derives \rn{Score} by joining the three relations on their shared variables $p$ and $t$: \rn{Treat} links each patient $p$ to the treatment $t$ it received and carries no embedding. \rn{Queries} and \rn{Keys} attach a query vector $q$ to each patient and a key vector $k$ to each treatment.
The join brings $q$ and $k$ together on every linked pair $(p,t)$, and the head records their product $q*k$ as that pair's embedding. The second rule joins \rn{Score} with the treatment values \rn{Values} on the shared attribute $t$, weighting each value $v$ by its score $a$. Its head, \rn{Attention}, keeps only the patient $p$ and drops $t$. 
This projection groups the treatments by patient and applies \fn{sum}, aggregating the weighted values $a*v$ to a single embedding per patient.
%\footnote{RelaNN also supports softmax via a function definition.}
%\sa{Explain a little more how each element is reflected in the example. This is the first example the reader sees - it should be very clear.}
%% --- yl: original example below used the outdated <...> embedding syntax. ---
%% --- Replaced with the actual R(content ; embedding) syntax and the :- neck (see Section 4). ---
% \begin{align*}
% \textsc{QK}(p,t)\angs{\mathsf{softmax}(q*k)}
% \leftarrow\, &
% \textsc{Treatment}(p,t),
% \\
% &\textsc{Queries}(p)\angs{q},\textsc{Keys}(t)\angs{k}\\
% %
% \textsc{Attention}(p)\angs{\mathsf{sum}(x*v)}
% \leftarrow\,  &
% \textsc{QK}(p,t)\angs{x},
% \textsc{Values}(t)\angs{v}
% \end{align*}
% \yl{Written as equations, the attention step is}
% \[
%   \mathrm{Score}(p,t) = q_p * k_t,
%   \qquad
%   \mathrm{Attention}(p) = \sum\nolimits_t \mathrm{Score}(p,t) * v_t ,
% \]
% \yl{which RelaNN expresses as the two rules}
% \begin{code}
% \rn{Score}(\ca{p},\ca{t}; \ea{q}*\ea{k}) \neck \rn{Treat}(\ca{p},\ca{t}), \rn{Queries}(\ca{p}; \ea{q}), \rn{Keys}(\ca{t}; \ea{k}) .
% \rn{Attention}(\ca{p}; \fn{sum}(\ea{a}*\ea{v})) \neck \rn{Score}(\ca{p},\ca{t}; \ea{a}), \rn{Values}(\ca{t}; \ea{v}) .
% \end{code}
%\yl{do names in example that are short so it will be one line. the task is to say this is our language, with example that it has both embedding and relatioins. no need for the latex equestions. no equations so it will be on dbs not on graphs.}
%\dl{No one knows up to now what a graph transformer is, if you show the latex and then the code, then it might make sense as a teaser}.
The program maintains an \e{assignment} of values to its parameters, and \e{training} adjusts them by minimizing a loss function.

Our approach adopts the idea of representing every database fact in two domains---the \e{logical domain} having the traditional database context, and the \e{numerical domain} that associates a meaning in the Euclidean space and allows for numerical reasoning, as studied in prior work on tuple embedding~\cite{DBLP:conf/icde/TonshoffFGK23,DBLP:conf/sigmod/CappuzzoPT20,DBLP:conf/sigmod/BordawekarS17,DBLP:conf/cikm/LubarskyTGK23,DBLP:conf/sigmod/MudgalLRDPKDAR18,DBLP:journals/ml/CvetkovIlievAV23}. This is
contrasted with existing in-database ML frameworks that typically bridge the two domains by integrating tensor data types and linear algebra operators into the query engine~\cite{li2017mlog,schule2019mlearn,Schle2019ML2SQLC,kunft2019lara,db4ml2020,li2024gaussml}, as also endorsed by the recent Tensor Logic~\cite{domingos2025tensorlogic} that unifies neural and symbolic AI by treating relations as sparse Boolean tensors, interpreting joins as tensor products and projections as sums over dropped dimensions. 
These frameworks require the user to explicitly define and maintain the connection between the logical and the vector representations, as well as their evolution through the composition of deep models.
%specify the representations, manually coordinating how relational attributes align with tensor matrices.
%In contrast, in this paper, every tuple is natively associated with an embedding; the classical relation and its embeddings are treated as a unified pair. Under this provenance-semirings inspired paradigm, %Similarly to the abstraction of database provenance as semirings~\cite{DBLP:conf/pods/GreenKT07}, where natural joins map to multiplication and projections/unions map to addition, 
%joins dictate embedding concatenation,
%while projections and unions are mapped to differentiable aggregation of the embeddings.
%dictate their additive reduction, merging the embeddings of collapsed tuples through differentiable aggregation functions. %[while projections and unions naturally govern their differentiable aggregation (via functions like sum, mean, or max).] 
%However, while Tensor Logic reduces all symbolic structures entirely to tensor equations, RelaNN maintains the classical relational model alongside continuous, trainable embeddings. 
%RelaNN utilizes database query engines to efficiently compute and cache the relational join graph while propagating and optimizing continuous, high-dimensional embeddings directly over this cached structure.

%In fact, the theoretical foundations of this approach have been recently studied in the form of \e{homomorphisms} (results) of patterns (join queries) in the graph domain~\cite{maehara2024deep,DBLP:conf/iclr/BaoJBCL25}.

\begin{figure}[t]
\centering
\resizebox{\columnwidth}{!}{%
\begin{tikzpicture}[
    stage/.style={rectangle, draw, rounded corners=2pt, minimum height=0.75cm, inner ysep=1pt, minimum width=1.4cm, align=center, font=\footnotesize},
    dbnode/.style={cylinder, draw, shape border rotate=90, aspect=0.25, minimum height=0.48cm, minimum width=0.8cm, align=center, font=\scriptsize, fill=blue!8},
    libnode/.style={rectangle, draw, rounded corners=6pt, minimum height=0.6cm, align=center, font=\scriptsize, fill=green!8},
    analogy/.style={font=\scriptsize\itshape, text=gray!70, align=center},
    arr/.style={->, thick, >=stealth},
    dblabel/.style={font=\scriptsize, align=center}
]
% torch.nn library (above Term Graph, as a compile-time input)
\node[libnode] (torchnn) at (6.0,1.04) {\texttt{torch.nn}};

% Database (above PyTorch Module, as an execution-time input)
\node[dbnode] (db) at (7.8,0.94) {Database};

% Pipeline stages (primary row)
\node[stage, fill=orange!10] (dsl) at (2.4,0) {RelaNN\\Program};
\node[stage, fill=orange!10] (era) at (4.2,0) {NRA};
\node[stage, fill=orange!10] (tg)  at (6.0,0) {Term\\Graph};
\node[stage, fill=orange!10] (pt)  at (7.8,0) {SQL + cuDF\\+ PyTorch};

% SQL analogy labels (bottom row, same x as stages for straight connectors)
\node[analogy, anchor=east] (aanchor) at (2.1,-0.9) {SQL\\analog:};
\node[analogy] (a1) at (2.65,-0.9) {SQL\\query};
\node[analogy] (a2) at (4.2,-0.9) {rel.\\algebra};
\node[analogy] (a3) at (6.0,-0.9) {logical\\plan};
\node[analogy] (a4) at (7.8,-0.9) {physical\\plan};

% Thin connectors from pipeline boxes down to analogy labels
\draw[gray!50, thin] ([xshift=0.25cm]dsl.south) -- (a1);
\draw[gray!50, thin] (era) -- (a2);
\draw[gray!50, thin] (tg)  -- (a3);
\draw[gray!50, thin] (pt)  -- (a4);

% Arrows: pipeline
\draw[arr] (dsl) -- (era);
\draw[arr] (era) -- (tg);
\draw[arr] (tg)  -- (pt);

% Compile-time / execution-time inputs (now from above)
\draw[arr, dashed] (torchnn) -- (tg);
\draw[arr, dashed] (db) -- node[dblabel, left, xshift=-1pt] {\textit{SQL data loader}} (pt);
\end{tikzpicture}%
}
\vskip-1em
\caption{Compilation pipeline of RelaNN. A program is compiled into Neuro-Relational Algebra (NRA), then to a term graph and a physical plan that executes over SQL, cuDF, and PyTorch. Neural operators (e.g., \texttt{Linear} and \texttt{ReLU}) are resolved at compile time from the torch.nn library. 
%At execution time, data loaders compile to SQL that runs in the database, connecting its stored relations to the physical plan.
}
\label{fig:pipeline}
%\bk{The top (gray part) looks like it is what you compile with. You want to emphasize that these are the pure relational parallels.}
\vskip-1em
\end{figure}
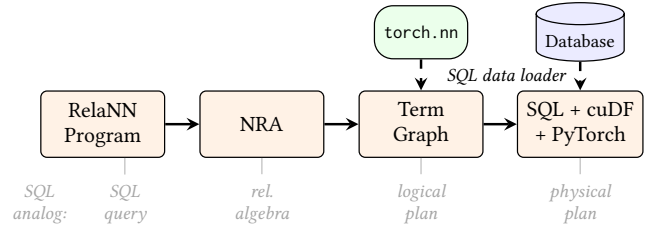

The theoretical foundations of this approach have recently been studied in terms of \e{homomorphisms} of patterns (join queries) in the graph domain~\cite{maehara2024deep,DBLP:conf/iclr/BaoJBCL25}. However, a key question remains open: whether and how such an approach can be realized in practice. This paper addresses this challenge. We introduce \e{RelaNN}\footnote{Source code: \url{https://github.com/yuvallu/relann}.}, an open-source system that enables the design of neural models directly over relational data through a declarative query language as described above. As illustrated in \Cref{fig:pipeline}, programs are first translated into Neuro-Relational Algebra (NRA) to incorporate embedding manipulation, and subsequently compiled into a physical plan across PyTorch, cuDF, and SQL to execute the model.
%programs are translated into Neuro-Relational Algebra (NRA) that incorporates embedding manipulation, and are then compiled into a physical plan over PyTorch, cuDF, and SQL that executes the resulting model.

NRA extends classic relational algebra by equipping every operator with two complementary semantics: the standard \e{content semantics}, determining how tuples are derived, and an \e{embedding semantics}, specifying how their embeddings are computed. In particular, joins combine tuples in the usual sense while concatenating their embeddings; projection and union are unified into a \e{projected union} operator that aggregates embeddings over multisets of tuples using aggregate functions such as sum, mean, or max; and transformation operators apply differentiable maps (e.g., feed-forward networks) to embeddings, introducing learnable parameters. To support learning, NRA is coupled with a \e{fit} operation that assigns values to these parameters by minimizing a loss function defined over the output of an NRA expression. Thus, %As a result, 
NRA expressions define end-to-end differentiable computations over relational data, unifying relational query processing with neural model design and optimization within a single formal framework.

Our experiments demonstrate several key properties of our approach. First, RelaNN can express a wide range of graph-learning architectures in concise and simple programs, including graph convolutional networks (GCN)~\cite{kipf2017semi}, relational GCN (R-GCN)~\cite{schlichtkrull2018modeling}, heterogeneous graph transformers (HGT)~\cite{hu2020heterogeneous}, hypergraph neural networks (HyGNN)~\cite{10184559}, and deep homomorphism networks (DHN)~\cite{maehara2024deep}.
%\dl{Here its importnat to differentiate what can be done in dgl pyg etc, and what goes beyond them due to the relational model being more powerful than just graphs.} \yl{ignore. no need here}
Across these models, RelaNN programs are significantly shorter than the reference implementations and closely mirror the mathematical formulations in the original papers. Second, RelaNN achieves comparable predictive accuracy to the corresponding PyTorch and PyG implementations, and remains within the same order of magnitude in runtime; in the case of DHNs, RelaNN achieves a considerable improvement of runtime, up to 14$\times$ faster. Third, RelaNN enables rapid design and iteration of task-specific models over relational data, where minor, localized programmatic changes 
%small, localized changes in the program 
lead to meaningful improvements in predictive performance, free of engineering overhead.

\section{Neuro-Relational Algebra (NRA)}
\label{sec:ff}
This section presents NRA---the formal framework that serves as the basis for the RelaNN language. NRA extends standard relational algebra with numerical updates to tuple annotations, which are defined in \Cref{sec:data-model}. 
%It is built on a relational database equipped with embedded annotations on tuples, defined in Section~\ref{sec:data-model}. 
\Cref{sec:nra} defines the NRA operators. Finally, \Cref{sec:ff-train} formalizes training as a loss-minimizing assignment of values to the parameters of an NRA expression.

\subsection{Neuro-Relational Data Model (NRM)}
\label{sec:data-model}
Our database and schema models extend the classic relational model~\cite{DBLP:journals/cacm/Codd70} with vector annotations of tuples, which we name \e{embeddings}. A schema $\scs$ associates a \e{content arity} $k$ and an \e{embedding dimension} $d$ with each relation name $R$. % \yl{Removed the $R/k\angs{d}$ schema notation and the open syntax question; stated in prose instead.}
% Hence, we assume that the schema $\scs$ is a collection of expressions of the form $R/k\angs{d}$ with unique relation names $R$ (i.e., no two $R/k\angs{d}$ have the same $R$).
We assume that relation names are unique within a schema.

% A \e{database} $D$ over the schema $\scs$ contains, for each such $R/k\angs{d}$, an \e{embedded relation} $(r,\eta)$, where $r$ is a relation with $k$ attributes and $\eta$ is an \e{annotation} that maps every tuple $t\in r$ into an \e{embedding} $\eta(t)\in\reals^d$.
A \e{database} $D$ over the schema $\scs$ contains, for each relation name $R$ in $\scs$ with content arity $k$ and embedding dimension $d$, an \e{embedded relation} $(r,\eta)$, where $r$ is a relation with $k$ attributes and $\eta$ is an \e{annotation} that maps every tuple $t\in r$ to an \e{embedding} $\eta(t)\in\reals^d$. We call $r$ the \e{content} of the embedded relation.

% In Section~\ref{sec:nra} we define NRA operators that may introduce \e{parameters} to the annotations of the tuples in the database $D$. In this case, we have a finite set $P$ of numeric parameters, and an assignment $\gamma:P\rightarrow\reals$ (that can be fit to training data, as explained in Section~\ref{sec:ff-train}). The annotation $\eta$ of each relation $r$ becomes fully specified only with respect to the assignment $\gamma$ to its parameters, and it then becomes an ordinate annotation that we denote by $\eta_{P\gets \gamma}$. We call such a database a \e{parametric database}, and we denote it as $(D,P,\gamma)$.

\subsection{Neuro-Relational Operators}
\label{sec:nra}

\def\oppara#1{\medskip\par\noindent\underline{#1}.\,}
\def\bvec#1{\mathord{\mathbf{#1}}}

Having fixed the data model in \Cref{sec:data-model}, we now define \e{Neuro-Relational Algebra} (NRA): a relational algebra whose objects are embedded relations. NRA operators take a finite number of embedded relations and return a single embedded relation. A finite composition of these operators is named an \e{NRA expression}. For each operator, we define its \e{content semantics} and its \e{embedding semantics}. The former treats the content of the derived embedded relation, while the latter treats its embeddings.

\oppara{Join ($\join$)}
This operator executes a natural join, pairing tuples from the given embedded relations $(r_1,\eta_1)$ and $(r_2,\eta_2)$ that agree on all shared attributes.
%This operator performs a natural join. It pairs tuples that agree on shared attributes over the content of two given embedded relations $(r_1,\eta_1)$ and $(r_2,\eta_2)$.
Let 
%Let the embedded relation 
$(r,\eta)\df(r_1,\eta_1)\join(r_2,\eta_2)$ be the join of the two embedded relations. 
The content semantics of join are defined as usual: 
%As for the content semantics, we define 
$r=r_1 \join r_2$. The embedding of each output tuple is formed by concatenating the contributing embeddings from the input relations. Hence, for $t\in r$:
\[
\eta(t)\df\eta_1(t[\bvec{A_1}])\concat\eta_2(t[\bvec{A_2}])
\]
where $\concat$ denotes concatenation and $\bvec{A_1}$ and $\bvec{A_2}$ are the attribute sequences of $r_1$ and $r_2$, respectively. 
%The output embedding has dimension $d_1+d_2$.

\oppara{Projected union ($\cup_{\alpha,\bvec{A}}$)}
This operator is a set-semantics $k$-ary union combined with a projection. It first projects the tuples of the $k$ given relations onto a common attribute sequence, and then aggregates the embeddings of tuples with identical content. Let $\bvec{A}$ denote the attribute sequence, and let $\alpha$ denote a multiset aggregator.
%We denote the attributes sequence $\bvec{A}$ and the multiset aggregator $\alpha$. 
Let $(r_1,\eta_1), \dots, (r_k,\eta_k)$ be $k$ embedded relations over the attributes $\bvec{A}$, and denote  $(r,\eta)\df\cup_{\alpha,\bvec{A}}\big((r_1,\eta_1),\dots,(r_k,\eta_k)\big)$. The projections of the incoming tuples onto $\bvec{A}$ are unified, that is, $r=\pi_{\bvec{A}}(r_1)\cup\dots\cup\pi_{\bvec{A}}(r_k)$. Each tuple $t\in r$ is associated with a multiset $B_t$ of input tuples whose projection over $\bvec{A}$ yields $t$, namely $B_t\df\bag{t'\in\cup_{i=1}^kr_i\ |\ t'|_{\bvec{A}}=t}$. The embedding of each $t\in r$ is then given by aggregating the embeddings of all the tuples from $B_t$:
\[
\eta(t)\df\alpha\bigl(\lmset \eta_i(t')\ |\  t'\in B_t \cap r_i,\ i\in\{1,\dots,k\}\rmset\bigr)\
\]
% \sa{I think we can say \[
% \eta(t)\df\alpha\bigl(\lmset \eta_i(t')\ |\  t'\in B_t\rmset\bigr)\
% \]
% because being in $B_t$ implies being in the union of $r_i$.
% }

Note that this operation generalizes union and projection operators. For $k=1$ we get the standard projection operator. For $k=2$ and identical relational schemas $\bvec{A}$ of $r_1$ and $r_2$, we get the union operator. The reason for this unification of operators is rooted in the embedding calculation. %\sa{from here until the end of the paragraph it is unclear. What does the assumption have to do with the unification of the two operations?} 
Any of the sub-operations along the way (projection or union of a subset of the $k$ given relations) may trigger the need for aggregation, due to duplicate tuples. Since we do not assume decomposability of the aggregation (e.g., mean), it must be applied only once, at the end of the full operation.
%If we want to allow an aggregation of both origin tuples received from a projection operation and a union operation, we have to execute the aggregation once. Otherwise, we are assuming the aggregation be decomposable, which is not necessarily the case (i.e. mean). For this reason we also don't strict the arity to be 2 as the other operators.

% The only assumption we make about the aggregator $\alpha$ is that it operates over multisets. In particular, it is not necessarily possible to break a projected union into an expression that comprises only simple projections and binary unions.

%It is not necessarily decomposable, meaning that it might not be possible to break down the aggregation computation into the underlying submultisets. For example, $\alpha$ might be the average aggregator. The average of $k$ numbers cannot be broken down into $k-1$ binary averages. In practice, we need to be able to calculate such aggregations, so we introduced a unified operator for all the operators that require embedding aggregation.

\oppara{Transformation ($\mathsf{T}_\tau$)} 
A transformation operator alters the embeddings %performs an alteration over the embeddings 
of a single embedded relation. Content is unchanged; a differentiable transformation $\tau$, composed of learnable layers and activations, is applied independently to each tuple embedding. %, which is where learnable layers and activations live. 
%Because learnable layers are acceptable in this operator, new parameters might be introduced in it. 
This is the only operator that may introduce new parameters, as part of the learnable layers.
We denote by $P_{\tau}$ the set of parameters associated with a transformation operator $\mathsf{T}_\tau$.
Formally, given an embedded relation $(r_0,\eta_0)$ and a transformation $\tau$ over its embeddings, the outcome $(r,\eta)\df\mathsf{T}_\tau(r_0,\eta_0)$ is given by $r\df r_0$ and $\forall t\in r: \eta(t)\df\tau(\eta_0(t))$.

%Note that the operation $\tau$ depends on the values assigned to the parameter set $P_{\tau}$. The instantiation of them is explained in Section~\ref{sec:ff-train}. %This is the only operator that introduces parameters.

% The embedding dimension may change due to the transformation. When the transformation operator calculated, values for the parameters must be in place. They are read through the assignment $\gamma$. It is the only NRA operator that consumes parameters from $P$ via the assignment $\gamma$; every other operator is parameter-free.

%\oppara{Other operators} NRA also includes the other operators that are not affected by the embedding, including selection ($\sigma$), difference ($\setminus$), and renaming ($\rho$).
\oppara{Other operators} NRA also includes selection ($\sigma$), set difference ($\setminus$), and renaming ($\rho$), which behave as in classical relational algebra and apply no transformation to embeddings: each output tuple keeps its input embedding.

\subsection{Training (fit)}
\label{sec:ff-train}
To evaluate an NRA expression, the parameters introduced by its transformation operators must be assigned values. Given an NRA expression $\phi$, we denote by $P_{\phi}$ the union of the parameter sets $P_\tau$ for each transformation $\tau$ in the composition of $\phi$. We now define a training operation that produces the best such values. %called %\sa{\e{fit}}.
%\e{training}

% To support the parametric updates we introduce a \e{meta-operation} which sits on top of NRA and together exhaust how $P$ is consumed and updated. It acts as a fixed optimization process similar to classic neural networks.

% For enabling the optimization process we require several objects to be defined: a target embedded relation $T\in D$ that contains the labels of a learning task. An NRA expression $e$, with its outcome having the same schema as $T$. A differentiable loss~$L$, and hyperparameters $H$ (learning rate, optimizer, number of epochs, $\ldots$). 

A \e{fit operation} accepts an NRA expression $\phi$, its parameter set $P_\phi$, and a database $D$. It returns an \e{assignment} $\gamma:P_\phi\to\mathbb{R}$ to the parameters, ideally one that minimizes a predefined optimization objective. In Section~\ref{sec:fit-statements} we lift this operation into a language-level statement that maintains $\gamma$ across calls.

\section{The RelaNN Language}
\label{sec:RelaNN_lang}

\begin{comment}
Take-home message:
- 
- 
OUTLINE:
- Rule
- TransformDef
- FunctionDef
- fit/pred?
- Example - simple
\end{comment}

% We summarize the concrete syntax of RelaNN programs as they are written, complementary to the formal framewok account in Section~\ref{sec:nra}. We detail a list of extensions to the basic model \bb{or how to call it?} that are vital to make RelaNN operational.

% \subsection{NRP} 

% Similarly to NRP definition, a \e{program} is a finite sequence of top-level statements, each terminated by a period. The main type of statements is a rule. Similarly to the base rules in the formal framework \bb{or how to call it? NRA?}, each rule statement extends the existing embedded relation by generating a new single embedded relation according to the specified conventions. The language expresses join rules, transformation rules, and aggregation rules using a unified \emph{single} rule form:

% \def\subin{_{\mathsf{in}}}
% \def\subout{_{\mathsf{out}}}

% An NRP takes as input a database $D\subin$ over a fixed schema $\scs$, and builds a parametric database $(D\subout,P,\gamma)$. Syntactically, an NRP is a sequence $\Pi = (\rho_1, \dots, \rho_n)$ of \e{rules}, each adding a new embedded relation, possibly with new parameters, to $(D\subout,P,\gamma)$. We consider three \bb{four? 3 primitive/basic and one optimization/special?} kinds of rules.

\def\subin{_{\mathsf{in}}}
\def\subout{_{\mathsf{out}}}
We next describe
%We summarize 
the concrete syntax of the RelaNN language, % as it is written, 
extending the Neuro-Relational Algebra introduced in Section~\ref{sec:ff}. 
%\sa{I think we can remove the next sentence:} We detail the building blocks of RelaNN that are vital to make it operational.

A \e{neuro-relational program (NRP)} is a finite sequence of \e{statements}. It takes as input a database $D\subin$ over a schema $\scs$ and produces an output database $D\subout$. 
%We call relations in $D\subin$ \e{extrinsic} or "database relations" (stored in the database) and relations added to $D\subout$ by program statements \e{intrinsic} (derived by the program). 
The core building blocks are \e{rule statements}, each adding a new embedded relation to $D\subout$ (Section~\ref{sec:rules}). As in Section~\ref{sec:ff}, rules may introduce parameters; the full set $P$, with an assignment $\gamma:P\to\mathbb{R}$, is held statefully alongside $D\subout$. 
%This extends the formal definition of a fit operation (Section~\ref{sec:ff-train}), which is stateless. 
\e{Fit statements} (Section~\ref{sec:fit-statements}) update the parameter assignment to fit data. \e{Predict statements} (Section~\ref{sec:predict-statements}) evaluate the program at the current assignment without updating it.

On top of these core statements, RelaNN provides two organizational primitives: \e{aliases} (Section~\ref{sec:aliasing}) for parameter sharing across rules, and \e{function definitions} (Section~\ref{sec:functions}) for code reuse.

\subsection{Rule Statements}
\label{sec:rules}

A RelaNN \e{rule statement} (or \e{rule} for short) is the basic component of an NRP. It takes as input a set of embedded relations, either from $D\subin$ or created by previous rules, and describes how to translate them
%It takes as input multiple embedded relations that already exist in the input database, or are created by previous rules. It describes how those embedded relations are translated 
into a new embedded relation that is introduced to the database. Each rule is directly compiled into an NRA expression, composed of neuro-relational operators (\Cref{sec:nra}). %the operators introduced in Section~\ref{sec:nra}. 
%\sa{each rule is compiled to an NRA expression?} 
The embedded relations in a rule take the form $R(\vec{x}; z)$, where $R$ is the relation name, and $\vec{x}$ and $z$ tuple-wise bind the relation's content variables and embeddings, respectively. 
RelaNN supports
%We distinguish 
two types of rules: \e{join rules} and \e{union rules}. Together with filter clauses in their body, they expose the join, union, projection, transformation, and selection operators of NRA (Section~\ref{sec:nra}).
%;set difference and rename do not have surface forms.

%\subsubsection*{Join rules}
%\label{sec:join-rules}
A \e{join rule} is an expression of the form
\[
R(\vec{x};\, \alpha(\tau)) 
\rneck 
R_1(\vec{x}_1; z_1), R_2(\vec{x}_2; z_2)\ , \cdots,\ R_k(\vec{x}_k; z_k),\; \varphi\ .
\]
It compiles into the NRA $\cup_{\alpha,\vec{x}}(\mathsf{T}_\tau(\sigma(R_1\join R_2\join\dots\join R_k)))$. First, the input embedded relations $R_1,\dots, R_k$ are joined on their shared variables using $k-1$ NRA join operations. This results in each tuple containing all the input variables $\vec{x_i}$, and the input embeddings concatenated together. If the (optional) \e{filter expression} $\varphi$ is specified, a selection operator is applied to the result. %A selection operator is applied on the result using the \e{filter expression} $\varphi$. The term $\varphi$ is optional, and if not mentioned it makes the selection operator ineffective. 
Next, an NRA transformation operator 
$\tau(z_1\oplus \dots \oplus z_k)$
% $\tau(z_1\cdot$$ \dots $$\cdot z_k)$ 
is evaluated. 
%The term that is used to express $\tau$ is called the \e{transformation expression}. \sa{what term? in the NRA or in the NRP rule? Where is this name (transformation expression) used - why is it defined here?}
The syntactic expression that specifies $\tau$ in the rule is called the \e{transformation expression}.
Finally, an NRA projected union operation is executed over $\vec{x}$ using the aggregator described in the \e{aggregation expression} $\alpha$. This yields the output embedded relation $R$. An example of a join rule can be seen in lines 3--4 of Example~\ref{ex:nrp_example}.
%\Cref{sec:nrp-example}, \sa{in lines 3-4 of the NRP}. \sa{consider labeling the NRP example from \Cref{sec:nrp-example} and referencing it here directly.}

%\subsubsection*{Union rules}
%\label{sec:union-rules}
A \e{union rule} is an expression of the form
\[
R(\vec{x};\, \alpha(\tau)) 
\rneck 
R_1(\vec{x}_1; z_1)\ |\ R_2(\vec{x}_2; z_2)\ |\ \cdots |\ R_k(\vec{x}_k; z_k),\; \varphi\ .
\]

In this rule all the embedded relations $R$ and $R_i$ must share the same structure, including the content attributes %schema
(i.e., $\vec{x}=\vec{x_i}$ for all $i$) and the embedding shape. It is compiled into the NRA expression $(\mathsf{T}_\tau(\sigma(\cup_{\alpha,\vec{x}}(R_1,R_2,\dots,R_k)))$.
The tuples are first unified using a projected union operation (which reduces to a plain union operator because a union rule requires all its relations to share the same content attributes). They are then optionally filtered by the selection operator using the filter expression $\varphi$. Finally, the embeddings are transformed using the transformation expression $\tau$.

% In this rule all the embedded relations $R$ and $R_i$ must share the same schema, including the content schema (i.e $\vec{x}=\vec{x_i}$ for all $i$), and the embedding shape. The tuples are first unified into a single embedded relation, with its embedding bound into a single embedding variable $z$. It then applies the transformation defined in the transformation expression $\tau(z)$, tuple-wise. The tuples are then projected into $\vec x$ and their embeddings are reduced by $\alpha$, resulting in the output embedded relation $R$. Notice that this rule applies an NRA expression composed of a transformation operator, followed by $k-1$ union operators and a projection operator. The body may also end with an optional filter $\varphi$ (Section~\ref{sec:filters}).

%\subsubsection*{Filter expressions.}
%\label{sec:filters}

The filter expression $\varphi$ is defined as a comma-separated list of atomic predicates:
$\varphi \df \varphi_1, \dots, \varphi_m$.
Each predicate $\varphi_i$ takes the form $e_1 \theta e_2$, where the operator $\theta \in \{=, \neq, <, \leq, >, \geq\}$, and $e_1, e_2$ are arithmetic terms over the body's content variables and constants. 
This list is interpreted conjunctively, meaning that its semantics correspond to the predicate $\varphi = \bigwedge_{i=1}^{m} \varphi_i$.

%RelaNN natively exposes only conjunction; there are no disjunction or negation operators in the rule body. Instead, a disjunctive guard $\varphi \vee \psi$ is expressed by writing a union rule over the embedded relations resulting from two separate rules that apply $\varphi$ and $\psi$ separately using a filter expression.

\subsection{Fit Statements}
\label{sec:fit-statements}
A \e{fit statement} realizes the fit operation (\Cref{sec:ff-train}) as a language construct. 
%While the abstract operation is stateless, the fit statement 
We maintain an assignment $\gamma$ to the parameters, and each invocation of the fir operation reads the current $\gamma$, computes a new assignment from the data, and writes it back.
Note that ``fit'' does not introduce new content; instead, it specifies how the parameters should be updated based on some given optimization hyperparameters. Syntactically, it is written as
\[
\texttt{?fit}\,\langle \text{kwargs} \rangle \quad \text{L}\ .
\]
% \[
% \texttt{?fit}\,\langle \text{kwargs} \rangle \quad L(;\,\ell) \rneck \text{Prediction}(\vec{x}), \text{Labels}(\vec{x})
% \]
$L$ plays the role of a loss function: % Its schema is constrained to be of arity 0 for content dimension and 1 for embedding dimension, meaning it's of the schema $L/0\angs{1}$.
its schema has the content arity 0 and embedding dimension 1. Therefore, it contains a single distinguished numeric cell, representing the loss value of an NRA expression. This value is then optimized over %according to 
the parameters derived from the NRA expression. The optimization process is configured via
%done according to 
optimization hyperparameters (e.g., learning rate, number of epochs, weight decay), which are specified in the keyword-argument list $\text{kwargs}$. Executing \texttt{?fit} updates the assignment $\gamma$ by minimizing the single embedding value exposed by $L$ (see Example~\ref{ex:nrp_example}). %An example appears in \Cref{sec:nrp-example} (line 10). \sa{same note about a direct reference.}

\subsection{Predict Statements}
\label{sec:predict-statements}

% NRPs might result with many different embedded relations, each introducing a different calculation path. Not all of them are required to be calculated given an NRP. For the matter of specifying a requested calculation path in the NRP, the predict statement is defined. A \e{predict statement} takes the role of executing an NRP. Given previous statements of an NRP, the predict statement calculates a specified embedded relation from one of the rules listed previously in the NRP. Syntactically, it is written as

NRP rule statements define an NRA expression, but never execute it. For %the sake of 
execution we introduce a \e{predict statement}. It takes the NRA expression that is defined by the NRP, the current assignment to the parameters (ideally, %that desirably were 
previously optimized using the fit statement), and a relation name to be evaluated. Syntactically, it is written as
\[
\texttt{?pred} \quad \text{R}\ .
\]
The embedded relation $R$ must be defined in
%exist in 
rules prior to the predict statement. After processing the predict statement, the embedded relation $R$ will be included %exist 
in the output database $D_{out}$. An example of a predict statement can be found in line 11 of Example~\ref{ex:nrp_example}. %\sa{same comment about direct reference. Looks like it's worth it since this is the third instance of referencing this example.}

% It is then fed into the embedded relation $\text{Output}$ which is now exposed concretely to the database. To understand the details of rules execution more in details refer to Section~\ref{sec:implementation}. Potentially a transformation $\tau$ which doesn't introduce new parameters might be evaluated as well. 

\subsection{Aliasing}
\label{sec:aliasing}
Programs may include \e{aliases} that introduce readable names for expressions reused across rules. An alias is a statement of the form
\[
\texttt{name} = e\,.
\]
where $e$ is an \e{alias expression}, of the same syntactic form as the transformation expression $\tau$ of a rule but evaluated \e{once, ahead of rule execution}, rather than per tuple and per rule. 

We distinguish two kinds of aliases by the form of $e$.
In a 
%A 
\e{scalar alias}, $e$ is %has $e$ as
an arithmetic expression over numeric constants, operators, and previously bound names (e.g.,\ $\texttt{d}=64$, $\texttt{h}=\texttt{d}/4$). It resolves to a numeric value.
A \e{parameterized alias} introduces new parameters into the parameter set $P$ (e.g.,\ $\texttt{p}=\texttt{Parameter}(64)$ or $\texttt{A}=\texttt{Linear}(2,3)$). Every textual occurrence of the aliased name in a rule's transformation expression refers to the same parameters, so the alias is RelaNN's mechanism for \e{parameter sharing} across rules. Examples appear in line~1 of Example~\ref{ex:nrp_example}.

\subsection{Function Definitions}
\label{sec:functions}
A \e{function definition} (or simply a \e{function}) abstracts a sequence of statements into a single named statement over embedded relations. Syntactically, it takes the form
\[
\texttt{def}\ F(R_1,\dots,R_n)\,:\;\; s_1\,.\ \dots\ s_m\,.\ \texttt{enddef}
\]
where $R_1,\dots,R_n$ are the \e{function parameters} standing for embedded relations supplied at each call, and the body $s_1,\dots,s_m$ is a non-empty sequence of statements. The last statement $s_m$ must be a rule, with its head determining the schema of the embedded relation returned by $F$. Intermediate statements are local to the function body.
A \e{call} to $F$ appears as a substitute for a relation name in a rule's body, written as
$
F(R_{a_1},\dots,R_{a_n})(\vec{x};\,z)
$
with argument relations supplied positionally in the first parenthesised list and the usual content attributes $\vec{x}$ and embedding variable $z$ in the second. Semantically, such a call is equivalent to inlining the body of $F$ with each parameter $R_i$ substituted by the actual argument $R_{a_i}$; the resulting embedded relation (produced by $s_m$'s head) is then exposed at the call site under $\vec{x}$ and $z$, like any other embedded relation. Functions therefore provide \e{code reuse}: the same body can be applied to different input relations, without code duplication. Functions are illustrated in Example~\ref{ex:nrp_example}, with lines 2--6 including a function definition, and line 7 including a call to a function.

% \subsubsection*{Function definitions}
% A \e{function definition} (or simply a function?) packages multiple following rules under one symbol:
% \[
% \texttt{def}\ F(R_1,\ldots,R_n)\ \ \cdots\ \texttt{enddef},
% \]
% where each $R_i$ is symbolic name of an embedded relation binded at calling. The body (should be introduced in the syntax? how?) is a sequence of rules and symbolic bindings; the \emph{last} rule’s head (LHS?) defines the schema and name of $F$’s output. Calls to $F$ appear as RHS atoms with argument relations supplied positionally.

\subsection{NRP Example}
\label{sec:nrp-example}
%This example demonstrates a basic NRP. % Consider a database with four embedded relations: \rn{Drivers}$/1\angs{16}$ with a learned embedding representing a driver, \rn{Races}$/1\angs{16}$ with  a learned embedding representing a race, \rn{Results}$/2\angs{16}$ with a learned embedding per result representing a result of a driver in a race, and \rn{Label}$/1\angs{4}$ with a supervised category label per driver (e.g. a demographic segment).
Consider a database with four embedded relations. \rn{Drivers} and \rn{Races} have content arity 1, each with a learned 16-dimensional embedding of a driver and of a race, respectively. \rn{Results} has content arity 2, with a learned 16-dimensional embedding of a driver's result in a race. \rn{Label} has content arity 1, with a 4-dimensional embedding holding a supervised category label per driver (e.g., a demographic segment). The following program computes, for each driver, a profile embedding aggregated from their results, and fits it to predict the driver's category.
\begin{listing}[t]
% Listing split into logical groups so each gap is a small \vspace, not a full blank
% line; firstnumber=last keeps the line numbering continuous across the groups.
\begin{code}[numbers=left]
\nm{d} = \nm{16}. \nm{k} = \nm{4}. \fn{Mix} = \fn{Linear}(\nm{2*d}, \nm{d}). \fn{Cls} = \fn{Linear}(\nm{d}, \nm{k}) .
\end{code}
\vspace{2pt}
\begin{code}[numbers=left,firstnumber=last]
\kw{def} \rn{DriverProfile}(\rn{Dr}, \rn{Ra}, \rn{Re}):
  \rn{Inter}(\ca{x},\ca{y}; \fn{Mix}(\fn{Concat}(\ea{z1}, \ea{z2})) + \ea{z3}) \neck
      \rn{Dr}(\ca{x}; \ea{z1}), \rn{Ra}(\ca{y}; \ea{z2}), \rn{Re}(\ca{x},\ca{y}; \ea{z3}) .
  \rn{Out}(\ca{x}; \fn{sum}(\ea{z})) \neck \rn{Inter}(\ca{x},\ca{y}; \ea{z}) .
\kw{enddef}
\end{code}
\vspace{2pt}
\begin{code}[numbers=left,firstnumber=last]
\rn{Profile}(\ca{x};\ea{z}) \neck \rn{DriverProfile}(\rn{Drivers},\rn{Races},\rn{Results})(\ca{x};\ea{z}) .
\rn{Loss}(;\,\fn{CrossEntropyLoss}()(\fn{Cls}(\ea{z_p}),\,\ea{z_l})) \neck
    \rn{Profile}(\ca{x}; \ea{z_p}),\ \rn{Label}(\ca{x}; \ea{z_l}) .
\end{code}
\vspace{2pt}
\begin{code}[numbers=left,firstnumber=last]
\texttt{?fit}\ \ensuremath{\langle} \nm{epochs}=\nm{100},\ \nm{lr}=\nm{0.01},\ \nm{weight\_decay}=\nm{5e-4} \ensuremath{\rangle}\ \rn{Loss} .
\texttt{?pred}\ \rn{Profile} .
\end{code}
\vspace{-1em}
\caption{NRP for driver category prediction.}
\label{ex:nrp_example}
\end{listing}
The first four statements are \e{aliases}: two scalar aliases fix the embedding width $d$ and the number of target classes $k$, and two parameterized aliases allocate the shared linear maps \fn{Mix} (pair-mixing) and \fn{Cls} (classifier head).

The function \rn{DriverProfile} takes \rn{Drivers}, \rn{Races}, and \rn{Results} relations as arguments. For each (driver, race, result) tuple it concatenates the driver and race embeddings, applies the shared linear map \fn{Mix}, and adds the result's embedding (rule \rn{Inter}); it then aggregates by driver with \fn{sum} (rule \rn{Out}) to produce one embedding per driver. The call on the following line binds this result to \rn{Profile}.

The \rn{Loss} relation calculates the final loss value using cross entropy over the previously bound linear transformation \fn{Cls} and \rn{Label} relation. It has empty content attributes (a single distinguished row), and its embedding expression reduces each joined pair to a scalar loss by passing $\ea{z_p}$ through the shared classifier \fn{Cls} and comparing the resulting logits against $\ea{z_l}$.

The final statements are fit and predict. The fit is executed over the embedded relation \rn{Loss}. It specifies the keyword-argument list 
\dslinline{\ensuremath{\langle}\nm{epochs}=\nm{100}, \nm{lr}=\nm{0.01}, \nm{weight\_decay}=\nm{5e-4}\ensuremath{\rangle}} which determines how the current assignment $\gamma$ of the parameter set $P$ should be updated. Crucially, $P$ here comprises \emph{both} the per-tuple embeddings stored in \rn{Drivers}, \rn{Races}, and \rn{Results}, \emph{and} the weights introduced by the aliases \fn{Mix} and \fn{Cls}; the fit statement updates them jointly so as to minimize the loss averaged over all (driver, driver's class, label) triples. The predict statement \dslinline{?pred Profile} then materializes \rn{Profile} in the output database under the updated $\gamma$.

% \paragraph{Example.}
% Suppose \texttt{Catalog(item; }$z_{\mathrm{cat}}$\texttt{)} (should be in formal definition of a schema? i.e \texttt{Catalog(1)<d>}?) stores one learned product embedding per \texttt{item}, and \texttt{Line(ord, item, qty; }$z_{\mathrm{line}}$\texttt{)} stores order lines.
% \begin{code}
% \nm{d} = \nm{8} .
% \kw{def} \rn{JoinCatLine}(\rn{Cat}, \rn{Ln}):
%   \rn{Out}(\ca{ord},\ca{item},\ca{qty}; \ea{z_cat} * \ea{z_line}) \neck \rn{Cat}(\ca{item}; \ea{z_cat}), \rn{Ln}(\ca{ord},\ca{item},\ca{qty}; \ea{z_line}) .
% \kw{enddef}
% \rn{LineExt}(\ca{ord},\ca{item},\ca{qty}; \ea{z}) \neck \rn{JoinCatLine}(\rn{Catalog}, \rn{Line})(\ca{ord},\ca{item},\ca{qty}; \ea{z}) .
% \rn{OrdTotal}(\ca{ord}; \fn{sum}(\ea{z})) \neck \rn{LineExt}(\ca{ord},\ca{item},\ca{qty}; \ea{z}) .
% \end{code}

% \bk{Use a different font for the examples. This is horrible for a paper, extremely wasteful with spacing... this is not a paper font.}

% \noindent
% \texttt{JoinCatLine} joins on \texttt{item} and combines only the two embeddings; \texttt{qty} is carried as relational data. The call builds \texttt{LineExt}; \texttt{sum} over line rows for each \texttt{ord} yields one embedding per order, in analogy to \texttt{GROUP BY ord}.

\subsection{Language Extensions}
\label{sec:lang-extensions}

\begin{comment}
    Take-home notes:
    - This is syntactic sugar
    - Its purpose is just to avoid repetition
    - We implemented compile time templates
    - Template syntax meaning + examples
    Not to talk about:
    - Materialization at compile time; weight sharing (same template+args)
    - We add symbolic variables to the symbol table, and only at runtime we apply an assignment to them.
\end{comment}
Beyond the core rule grammar of Section~\ref{sec:rules}, RelaNN provides two language extensions.
\e{Encoding} and \e{decoding} move values between a relation's content attributes and its embedding.
\e{Templates} factor out repeated statements and repeated body relations.

\subsubsection*{Encoding and decoding}
\label{sec:encode-decode}
%Real relational databases store content, not embeddings: their tables hold raw column values.
%The core rule grammar of Section~\ref{sec:rules} computes over embeddings, 
RelaNN provides machinery for producing initial embeddings from the database and writing back the results.
\e{Encoding} reads raw content columns into embeddings, giving a network its initial features.
\e{Decoding} writes a learned embedding back as a content attribute, so a prediction becomes a regular column of the database.
Both directions use a uniform $[\,\cdot\,]$ bracket syntax.

As an example, suppose the \rel{Drivers} table has a \ca{bio} column---the text of the Wikipedia page linked by each driver's \ca{url}---and we want to learn a per-driver score and store it as a new column.
The architecture \e{encodes} \ca{bio} into a feature embedding, runs a network over it, and \e{decodes} the learned score into a new \rel{Pred} relation:
\begin{code}
\rn{DriverFeat}(\ca{did}; [\fn{TextEncoder}(\ca{bio})]) \neck \rn{Drivers}(\ca{did},\ca{bio}) .
\cmt{// ... intermediate rules computing the score}
\rn{Pred}(\ca{did}, [\ea{c}]) \neck \rn{Score}(\ca{did}; \ea{c}) .
\end{code}
In the first rule, an \e{encoding} bracket inside the head's transformation expression reads the \ca{bio} attribute into the embedding.
Because \ca{bio} is text, a rich type, the bracket wraps an encoder \nrcode{nn.Module}, as in $[\fn{TextEncoder}(\ca{bio})]$.
Numeric and boolean attributes need no module and tensorize directly, written simply as $[\ca{col}]$, and multiple bracketed items concatenate along the last dimension.
In the last rule, a \e{decoding} bracket in the head's \emph{content} position runs the other way, writing an embedding value back as a content attribute: $[\ea{c}]$ decodes \rn{Score}'s learned embedding $\ea{c}$ into a column of \rn{Pred}.
%The engine fills this column after the forward pass, so the prediction becomes a regular relational attribute.

\subsubsection*{Templates}
\label{sec:templates}

Neural architectures over relational data often contain dozens of near-duplicate definitions, one per attention head, edge type, or layer.
Inspired by C++, \e{templates}
%,  templates~\cite{veldhuisen1998templates}, 
factor out this repetition: a \e{statement template} factors out repeated statements, and a \e{body template} factors out repeated body relations within a single rule.
We use eight-head attention as a running example, where each head transforms an input embedding through its own learned matrix and the head outputs are combined.
Written out in full, this example takes eight per-head rules and one rule that combines them:

\begin{code}
\rn{AttHead1}(\ca{k}; \fn{Linear}(\nm{d}, \nm{d})(\ea{z})) \neck \rn{Input}(\ca{k}; \ea{z}) .
\ensuremath{\smash[t]{\vdots}}
\rn{AttHead8}(\ca{k}; \fn{Linear}(\nm{d}, \nm{d})(\ea{z})) \neck \rn{Input}(\ca{k}; \ea{z}) .
\rn{MultiHead}(\ca{k}; \fn{Concat}(\ea{z1}, \cmt{...}, \ea{z8})) \neck
  \rn{AttHead1}(\ca{k}; \ea{z1}), \cmt{...}, \rn{AttHead8}(\ca{k}; \ea{z8}) .
\end{code}

\paragraph{Statement templates.}
A \e{statement template} parameterizes a statement by an index in angle brackets. For the running example, this collapses the eight rules into one:
\begin{code}
\rn{AttHead}\tp{<i>}(\ca{k}; \fn{Linear}(\nm{d}, \nm{d})(\ea{z})) \neck \rn{Input}(\ca{k}; \ea{z}) .
\end{code}
A template is a pattern---it produces nothing until a rule \e{invokes} it. When \rn{AttHead}\tp{<i>} is invoked, the compiler materializes a concrete \fn{Linear}(\nm{d},\nm{d}) for each distinct argument, before term-graph construction.
Arguments may range over schema type names, integers, or string labels.
Invocations with identical arguments share parameters; invocations with different arguments yield independent ones.

\paragraph{Body templates.}
The combine rule still lists eight body relations.
A \e{body template} replicates one body relation across an \e{iteration domain} in square brackets, either an integer range or the tuples of a relation.
The \repjoin{} replicator replicates the relation as a conjunction, collapsing the eight-way body to a single line:
\begin{code}
\rn{MultiHead}(\ca{k}; \fn{Concat}(*\ea{z})) \neck \rn{Att}\tp{<i>}(\ca{k}; \ea{z}) ,... [\nm{i} = \nm{1} to \nm{8}]
\end{code}
Each replica binds \nm{i} afresh, indexing the \rn{AttHead}\tp{<i>} statement template, and the splat $*\ea{z}$ gathers the eight head embeddings for \fn{Concat}.
Its disjunctive counterpart \repunion{} unions the replicas instead, so the rule head \e{adds} the heads with \fn{sum} rather than concatenating them.
Replicators expand at compile time into ordinary body relations, adding no runtime machinery.

\section{Implementation}
\begin{comment}
Take home message:
- We have open source
- Serious implementation of a system we can use.
- Inspired by databases.
OUTLINE:
- available as open source, tutorials, and demos, ready for use at @git..
- figure that explains how it looks like (just like DB's!).
- connect to db, IR (similar to translation to RA, we have NRA), given an NRA, we have a term graph, which is our logical query plan. Then we have the term graph to physical query plan (torch and cudf). Do a figure (inspired by SQL figures). For example, in brackets, put the SQL terminology.

- Pipeline overview (DSL to PyTorch; engine handles indices and differentiability)
- Subsections: From RelaNN to NRA, Term graph, Compile to PyTorch, Templated programs
\end{comment}

The RelaNN pipeline (Figure~\ref{fig:pipeline}) follows the architecture of a classic DBMS.
A RelaNN program (NRP, Section~\ref{sec:RelaNN_lang}) is translated statement-by-statement into an NRA expression, represented as a term graph (Section~\ref{sec:from-RelaNN-to-nra}). When a \texttt{fit} or \texttt{predict} statement is invoked, the subgraph that produces that statement's target relation is compiled to a \emph{physical plan} over cuDF~\cite{cudf} and PyTorch operations (Section~\ref{sec:compile-to-pytorch}) that execute joins, projections, and learned transformations end-to-end on GPU. Data loaders then connect the corresponding database relations to the physical plan via SQL queries.

RelaNN is a Python-embedded DSL parsed with Lark~\cite{lark}.
Transformation operators such as \texttt{Linear} and \texttt{ReLU} are resolved by name from the caller's Python namespace, so any user-defined \texttt{nn.Module} can be referenced directly in a rule.
The system is available as an open-source prototype with demos and documentation.
% \footnote{\url{https://github.com/yuvallu/relann}}

% As in a classical DBMS, this pipeline separates the user's specification of a neural architecture from how the system executes and optimizes it.
% %This separation enables the compiler to apply graph rewrites on the term graph (common-subexpression elimination, scope reduction and more) and to cache content-only indices across training iterations, reducing execution cost without changing the user's program (the Optimizations subsection).
% This separation between specification and execution is what makes systematic, query-style optimization possible, a direction we discuss in Section~\ref{sec:discussion} (Future Work). \dl{Just skip this paragrph, we dont need to say we will talk about something in chapter7 in chapter 5.}

\subsection{From RelaNN to NRA Term Graph}
\label{sec:from-RelaNN-to-nra}
\begin{figure}[t]
\centering
\begin{tikzpicture}[
    opnode/.style={rectangle, draw, rounded corners=2pt, minimum height=0.5cm, align=center, font=\footnotesize, inner sep=3pt},
    leafnode/.style={rectangle, draw, rounded corners=2pt, minimum height=0.5cm, minimum width=1.5cm, align=center, font=\footnotesize, fill=blue!12, inner sep=3pt},
    joinfill/.style={fill=orange!20},
    transfill/.style={fill=purple!15},
    aggfill/.style={fill=green!15},
    arr/.style={->, >=stealth, shorten >=1.5pt, shorten <=1.5pt}
]
\node[leafnode] (ue) at (-1.4, 0)   {\rel{Drivers}};
\node[leafnode] (me) at ( 1.4, 0)   {\rel{Results}};
\node[opnode, joinfill]  (jn) at (0, 0.72) {Join $\join$};
\node[opnode, transfill] (tr) at (0, 1.46)  {Transformation $\mathsf{T}_{\Linear(2d,d)\,\circ\,\Concat}$};
\node[opnode, aggfill]   (ag) at (0, 2.20) {Projected union $\cup_{\texttt{sum},\,x}$};
\node[inner sep=3pt, font=\footnotesize] (out) at (0, 2.92) {\rel{DriverAgg}};
\draw[arr] (ue) -- (jn);
\draw[arr] (me) -- (jn);
\draw[arr] (jn) -- (tr);
\draw[arr] (tr) -- (ag);
\draw[arr] (ag) -- (out);
\end{tikzpicture}
\caption{Term graph (logical plan) for \rel{DriverAgg}. Leaves are database relations and inner nodes are NRA operators.}
%: join, transformation, and projected union.}
\label{fig:term-graph}
\vskip-1em
\end{figure}
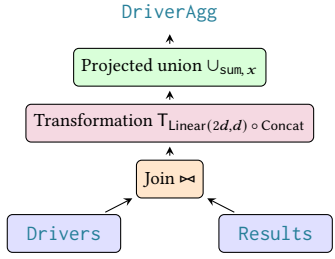

\begin{comment}
OUTLINE:
- Rule to term-graph fragment (nodes: join, transform, agg, project)
- RHS vs LHS (join structure; transformation + optional aggregation)
- Example (e.g. UserMovieRating)
\end{comment}
Each RelaNN rule translates to an NRA subexpression that captures its declarative semantics, as detailed in Section~\ref{sec:rules}.
% \bb{Not sure it's a problem, but we didn't define NRA subexpression} \yl{I think its a common terminology and we don't need to define sub-expression.}
For example, consider the rule \rel{DriverAgg}, which combines each driver's embedding with that driver's result embeddings (\Cref{sec:nrp-example})
\begin{code}
\rn{DriverAgg}(\ca{x}; \fn{sum}(\fn{Linear}(\nm{2*d},\nm{d})(\fn{Concat}(\ea{z1}, \ea{z2})))) \neck
    \rel{Drivers}(\ca{x}; \ea{z1}), \rel{Results}(\ca{x},\ca{y}; \ea{z2}) .
\end{code}
The corresponding NRA statement is
\begin{equation*}
\resizebox{0.92\columnwidth}{!}{$\rel{DriverAgg} \;=\; \cup_{\texttt{sum},\,x}\!\left(\mathsf{T}_{\Linear(2d,d)\,\circ\,\Concat}\!\left( \rel{Drivers} \join \rel{Results} \right)\right)$}\,.
\end{equation*}
%This NRA expression precisely captures the rule's declarative semantics, as defined in Section~\ref{sec:rules}: join the body relations, apply the head's transformation, and take the projected union on the head's keys.
The compiler assembles all NRA subexpressions of a program into a single \e{term graph}: a DAG of NRA operators that captures every rule's semantics.
Its inner nodes are NRA operators, its leaves reference the program's database relations, and its directed edges run from each operator to its consumers.
The \e{logical query plan} for a given \texttt{fit} or \texttt{predict} statement is then the subgraph of the term graph rooted at that statement's target relation.
Figure~\ref{fig:term-graph} illustrates the resulting term graph for \rel{DriverAgg}.
% The term graph can also be generated from real RelaNN code via scripts/gen_term_graph_fig.py.

% Dropped from §4.2 per Benny's comment below; preserved here in case we want
% to weave a phrase into §6 Discussion (Query-plan optimization paragraph).
% Compared to prior GNN compilers~\cite{xie2022graphiler,yu2023hector}, which
% also use DAG-based intermediate representations to capture message-passing
% patterns, the NRA term graph is schema-aware, encoding relational operations
% (joins, projections) alongside neural transformations.
% \bk{this is a strange statement. Do they operate over a relational database and ignore the schema? In what sense are they NOT schema aware? Also, is this a true difference? It seems like a minor difference from the previous compilations, accoridng to what you write here.}

\subsection{From Term Graph to Physical Plan}
\label{sec:compile-to-pytorch}
\begin{comment}
Take home notes:
- one of the things that makes RelaNN fast is that it is transpiles to torch, we support all the eco-system of torch. and the users enjoy all the optimizations and benefits of torch.
- 
OUTLINE:
- Every logical operator transforms to phisical operator.
- then we can do optimizations over the torch graph.
- it transpiles to cudf and torch.
- show examples of this mapping of each logical operator to phisical operator.
    - Scatter operations (torch_scatter; projection reducers)
    - same memory, cites papers, examples, eplenations.. ×œ×”×¨×‘×™×¥, torch graph cite blogs or websites... we used them in the implementation, we are seriouse.
    - example, join indices, aggregate...
- each operator is diff and backprop, so the entire torch graph is differentiability and backprop.
 % Question to Dean: Should we, instead of calling it instantiate (content pass), say we have smart content caching optimization? yes
 because in training we train update the weigths multiple times in the same batch on the same content, we actually implemented an optimizaiton that calculates it only once (without calling it "instantiate").
\end{comment}

The term graph defines what to compute; the physical plan defines how to execute it on a GPU.
Each NRA operator compiles to a \emph{subgraph} of physical operations.
Data loaders compile to an SQL subtree that runs in the underlying database.
All other operators compile to cuDF and PyTorch operations that maintain the alignment between content rows and embedding rows while preserving the gradient path through the embeddings.
The core node types compile as follows (others analogously):

\def\mypara#1{\smallskip\par\noindent\underline{#1}}

\mypara{Data loaders} compile to an SQL subtree that executes in the underlying database, returning the \emph{content} of an input relation as rows of a cuDF \nrcode{DataFrame}. For example, a selection or projection on a database relation can be pushed into this subtree, so the database filters and prunes columns before any content reaches the GPU.

% \bb{\mypara{Data loaders} compile to an SQL subtree that executes in the underlying database, returning the \emph{content} of an input relation as rows of a cuDF \nrcode{DataFrame}. The corresponding \emph{embedding} is supplied by the engine, either as a learnable per-tuple embedding or as the output of an encoding operator (Section~\ref{sec:encode-decode}), and is row-aligned with the loaded content. The content query could be leveraged to optimize data retrieval. For example, a selection or projection on a database relation can be pushed into the SQL subtree, so the database filters and prunes columns before any content reaches the GPU.}

%\dl{Just a thought, not sure about it. when doing rewriting to push down more to the sql, do we want to simply have the entier query be one node or have the term graph also have subtrees that are sql operators? It might clean up the implementation to talk about node/subtrees being sql.}
\mypara{Join} ($\join$) compiles to a cuDF \nrcode{DataFrame} merge that emits row-index tensors, which \texttt{torch.index\_select} uses to fetch embeddings with full autograd support.

\mypara{Selection} ($\sigma$) uses the same differentiable \texttt{torch.index\_select} with a boolean-mask index.

\mypara{Projected union} ($\cup_{\alpha,\bvec{A}}$) compiles to a cuDF groupby that emits a group-index tensor, over which \e{scatter operations}~\cite{torch_scatter} (\texttt{scatter\_add}, \texttt{scatter\_mean}, or \texttt{scatter\_max}, selected to match the aggregator~$\alpha$) apply the operator's multiset aggregation to the embeddings of each group. This is a sparse aggregation---it reduces variable-sized groups directly over the index tensor, without materializing a dense adjacency matrix---and gradients flow back through the scatter primitive. %\dl{The point here is not that pyg uses them, but that they are used to aggregate embeddings in a sparse way, without requiring large adjacency matrices or something, if you talked about bag semantics in the theory we can connect that here, but keep the commentaria consistent across bullets, we explain how the logical operatiosn translate to low level stuff in a way that makes it sounds technical. This comment about pyg is irrelevant. Maybe the what we should communicate across bullets is that the gradients are maintained using low level tensor and sparse tensor operations?}

\mypara{Transformation} ($\mathsf{T}_\tau$) nodes apply their \nrcode{nn.Module} (\texttt{nn.\allowbreak Linear}, \texttt{nn.\allowbreak LayerNorm}, or a custom layer) directly via a PyTorch forward call, so their parameters receive gradients automatically.

\smallskip

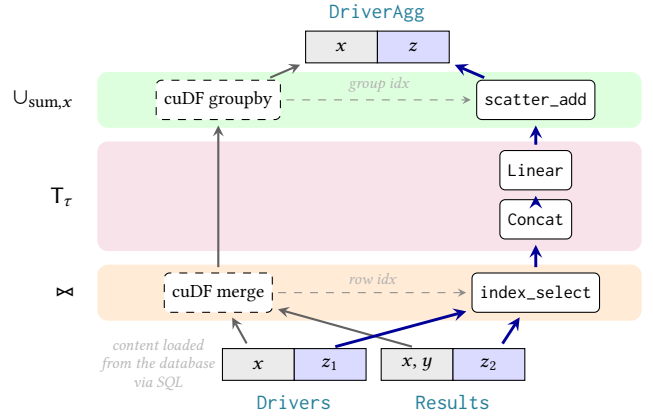
\begin{figure}[t]
\centering
\begin{tikzpicture}[
    cell/.style={rectangle, draw, minimum width=0.95cm, minimum height=0.42cm, align=center, font=\footnotesize, inner sep=1pt},
    embcell/.style={cell, fill=blue!14},
    concell/.style={cell, fill=black!8},
    cudfop/.style={rectangle, draw, dashed, rounded corners=2pt, minimum height=0.52cm, align=center, font=\footnotesize, inner sep=3pt, fill=white},
    torchop/.style={rectangle, draw, rounded corners=2pt, minimum height=0.52cm, align=center, font=\footnotesize, inner sep=3pt, fill=white},
    ername/.style={font=\footnotesize},
    bandlbl/.style={font=\normalsize, anchor=east},
    grad/.style={->, >=stealth, shorten >=2pt, shorten <=2pt, line width=1.1pt, color=blue!60!black},
    cont/.style={->, >=stealth, shorten >=2pt, shorten <=2pt, thick, color=black!62},
    idx/.style={->, >=stealth, shorten >=2pt, shorten <=2pt, dashed, color=black!45},
    edgelabel/.style={font=\scriptsize\itshape, text=gray!60}
]
% --- logical-operator bands (drawn first, behind everything) ---
\fill[orange!16, rounded corners=4pt] (-3.95,0.58) rectangle (3.25,1.32);
\fill[purple!11, rounded corners=4pt] (-3.95,1.50) rectangle (3.25,2.95);
\fill[green!13,  rounded corners=4pt] (-3.95,3.13) rectangle (3.25,3.87);
\node[bandlbl] at (-4.15,0.95) {$\join$};
\node[bandlbl] at (-4.15,2.22) {$\mathsf{T}_\tau$};
\node[bandlbl] at (-4.15,3.50) {$\cup_{\mathrm{sum},x}$};
% --- input embedded relations (content cell | embedding cell) ---
\node[concell] (dc) at (-1.82,0) {$x$};
\node[embcell] (de) at (-0.87,0) {$z_1$};
\node[ername] at (-1.35,-0.5) {\rel{Drivers}};
\node[concell] (rc) at (0.28,0) {$x,y$};
\node[embcell] (re) at (1.23,0) {$z_2$};
\node[ername] at (0.75,-0.5) {\rel{Results}};
% data loaders: the content of the input ERs is loaded from the database via SQL
\node[font=\scriptsize\itshape, text=gray!55, align=center] at (-3.15,0) {content loaded\\from the database\\via SQL};
% --- physical operators ---
\node[cudfop]  (mrg) at (-2.35,0.95) {cuDF merge};
\node[torchop] (gth) at (1.85,0.95) {\texttt{index\_select}};
\node[torchop] (cat) at (1.85,1.92)  {\texttt{Concat}};
\node[torchop] (lin) at (1.85,2.57)  {\texttt{Linear}};
\node[cudfop]  (grp) at (-2.35,3.50) {cuDF groupby};
\node[torchop] (sca) at (1.85,3.50) {\texttt{scatter\_add}};
% --- output embedded relation (content cell | embedding cell) ---
\node[concell] (oc) at (-0.72,4.20) {$x$};
\node[embcell] (oe) at (0.23,4.20) {$z$};
\node[ername] at (-0.25,4.65) {\rel{DriverAgg}};
% --- content edges (non-differentiable) ---
\draw[cont] (dc) -- (mrg);
\draw[cont] (rc.north) -- (mrg);
\draw[cont] (mrg) -- (grp);
\draw[cont] (grp) -- (oc);
% --- embedding edges (gradient-carrying) ---
\draw[grad] (de.north) -- (gth);
\draw[grad] (re) -- (gth);
\draw[grad] (gth) -- (cat);
\draw[grad] (cat) -- (lin);
\draw[grad] (lin) -- (sca);
\draw[grad] (sca) -- (oe);
% --- index tensors crossing from content track to embedding track ---
\draw[idx] (mrg) -- node[edgelabel,above]{row idx} (gth);
\draw[idx] (grp) -- node[edgelabel,above]{group idx} (sca);
\end{tikzpicture}
\vskip-0.5em
\caption{Physical plan for \rel{DriverAgg}. Each NRA operator 
%(shaded band, colored as in Figure~\ref{fig:term-graph}) 
expands into cuDF operations on content %(dashed boxes) 
and PyTorch operations on embeddings. Dashed arrows are non-differentiable index tensors, and gradients flow only along the blue chain.}
\label{fig:gradient-flow}
\vskip-1em
\end{figure}

\noindent Together, the subgraphs of the NRA operations form a single computation graph that \texttt{torch.autograd}~\cite{Paszke2017AutomaticDI} traces dynamically. Any \nrcode{nn.Module}, built-in or custom, can appear in a transformation and is trained end-to-end with the other operations.
Figure~\ref{fig:gradient-flow} shows the physical plan compiled from the \rel{DriverAgg} term graph.

\section{Experiments}
\label{sec:experiments}

We organize our evaluation around three questions.
First, can RelaNN express a wide range of published graph-learning architectures, from classical message passing, through graph transformers, to architectures that go beyond message passing, reproducing each model's reference accuracy in fewer lines of code (Section~\ref{sec:exp-known-archs})?
Second, how does the per-epoch runtime of a RelaNN program compare to a hand-written reference implementation (Section~\ref{sec:exp-runtime})?
Third, can a data owner use RelaNN to design a task-specific architecture over a multi-table database in a handful of rules, and improve it through small, local edits (Section~\ref{sec:exp-design})?
%Section~\ref{sec:exp-setup} outlines the experimental setup.

%% ===================================================================
\subsection{Experimental Setup}
\label{sec:exp-setup}
%% ===================================================================

\subsubsection*{Architectures and datasets.}
We evaluate RelaNN on five published architectures.
%\ylinline{TODO: inline itemize}
$\bullet$\, 
\emph{Standard GNN architectures:} GCN~\cite{kipf2017semi} and R-GCN~\cite{schlichtkrull2018modeling} cover standard and relational message passing. We compare against PyG~\cite{fey2019fast} for GCN and a newer implementation~\cite{thanapalasingam2022rgcn} for R-GCN, on the Cora and AIFB datasets respectively, both from the original papers.
$\bullet$\, 
\emph{Graph transformers:}
HGT~\cite{hu2020heterogeneous} is a heterogeneous graph transformer. We compare against both the original implementation and the newer PyG one, on the DBLP dataset~\cite{DBLP:conf/www/FuZMK20} from the PyG HGT tutorial.\footnote{\url{https://github.com/pyg-team/pytorch_geometric/blob/master/examples/hetero/hgt_dblp.py}}
$\bullet$\, 
\emph{Beyond GNNs:}
We also evaluate RelaNN on two architectures that go beyond the expressive power of GNNs, Deep Homomorphism Networks (DHNs)~\cite{maehara2024deep} and Hypergraph Neural Networks (HyGNNs)~\cite{10184559}. For both, we compare against the baselines and datasets from the original papers: the CSL and EXP expressivity benchmarks for DHNs, two synthetic datasets that ordinary message passing cannot solve, and the TWOSIDES and DrugBank drug-interaction datasets for HyGNNs.

\paragraph{Metrics.}
For each architecture, we report three metrics (Table~\ref{tab:unified-results}):
test-set accuracy, wall time of a single training epoch, and number of non-blank, non-comment lines of code. Accuracy and wall time are measured over five runs (seeds 42--46). Every baseline is a PyTorch implementation, the same framework RelaNN compiles to.
% Accuracy is the test-set metric from the original paper, with the mean and standard deviation over five runs (seeds 42--46).
% Runtime is the per-epoch training wall-clock time, reported as the median and standard deviation over the same runs.
% Code size is the number of non-blank, non-comment lines of the model definition---the architecture itself, not the data-loading, preprocessing, or training code---counted the same way for our program and the baseline.

\paragraph{System setup.}
We used a single Linux server (Ubuntu 22.04.5) with two Intel Xeon CPUs, 503\,GiB of RAM, and a single NVIDIA A40 GPU whose graphics and memory clocks were locked for stable timing.
Software versions are PyTorch 2.5.0 and CUDA 12.4.
% We enabled PyTorch's deterministic-algorithm mode and disabled cuDNN autotuning and nondeterministic kernels globally.
% For both RelaNN and the baselines, per-epoch time covers the training loop only, excluding one-time setup (such as compilation) and evaluation.
% All scripts, data loaders, and seeds are included in the RelaNN repository.

%% ===================================================================
\subsection{Implementing Known Architectures}
\label{sec:exp-known-archs}
%% ===================================================================

We implemented each of the five architectures as a RelaNN program.
RelaNN reproduces the reference accuracy while expressing each model in roughly $4\times$ fewer lines of code compared to the baseline (Table~\ref{tab:unified-results}). Beyond the model, a baseline also needs data-loading and training code, which a RelaNN reduces to loading its relations and a single \texttt{fit} call---widening the gap for a whole-program comparison.
\begin{table}[t]
\centering
\caption{%Five published
Architectures in RelaNN, each compared to its official reference (Section~\ref{sec:exp-setup}). Accuracy is the paper's test metric (\%, mean$_{\pm\text{std}}$), Time is per-epoch wall-clock time (ms, median$_{\pm\text{std}}$), and LOC counts non-blank, non-comment lines of the NN model definition. GCN and HGT run from shared initial weights, so their accuracies match by construction.}
\label{tab:unified-results}
\vskip-1em
\scriptsize
\setlength{\tabcolsep}{2pt}
\begin{tabular}{@{}llcccccc@{}}
\toprule
 & & \multicolumn{2}{c}{Accuracy (\%)} & \multicolumn{2}{c}{Time (ms/epoch)} & \multicolumn{2}{c}{LOC} \\
\cmidrule(lr){3-4}\cmidrule(lr){5-6}\cmidrule(lr){7-8}
Arch.\ & Dataset & RelaNN & Baseline & RelaNN & Baseline & RelaNN & Baseline \\
\midrule
GCN                      & Cora     & $80.1_{\pm 1.6}$  & $80.1_{\pm 1.6}$ & $6.8_{\pm 0.3}$    & $4.9_{\pm 0.4}$   & 21 & 217 \\
\midrule
HGT 1L$^a$               & DBLP     & $76.8_{\pm 2.3}$  & $76.8_{\pm 2.3}$ & $38.6_{\pm 4.4}$   & $34.8_{\pm 1.3}$  & 41 & 186 \\
HGT 2L$^a$               & DBLP     & $72.5_{\pm 3.8}$  & $72.5_{\pm 3.8}$ & $103.6_{\pm 3.5}$  & $52.5_{\pm 1.0}$  & 42 & 186 \\
pyHGT$^a$          & DBLP     & $79.5_{\pm 0.4}$  & $79.5_{\pm 0.4}$ & $40.4_{\pm 2.5}$   & $66.6_{\pm 1.3}$  & 43 & 143 \\
\midrule
HyGNN                    & TwoSides & $87.7_{\pm 0.2}$  & $87.8_{\pm 0.2}$ & $61.9_{\pm 0.9}$   & $73.5_{\pm 4.8}$  & 41 & 150 \\
HyGNN                    & DrugBank & $91.9_{\pm 0.5}$  & $89.9_{\pm 5.6}$ & $163.5_{\pm 0.6}$  & $482.5_{\pm 5.9}$ & 41 & 150 \\
\midrule
DHN (C2:4)$^b$           & CSL      & $22.0_{\pm 8.4}$  & $28.1_{\pm 0.9}$  & $45.9_{\pm 4.2}$   & $232.0_{\pm 8.7}$  & 21 & 259 \\
DHN (C2:4)$^b$           & EXP      & $50.0_{\pm 0.0}$  & $49.0_{\pm 0.6}$  & $45.8_{\pm 0.7}$   & $514.5_{\pm 12.8}$ & 21 & 259 \\
DHN (C2:10)$^b$          & CSL      & $100.0_{\pm 0.0}$ & $100.0_{\pm 0.0}$ & $275.8_{\pm 25.8}$ & $561.0_{\pm 4.0}$  & 76 & 259 \\
DHN (C2:10)$^b$          & EXP      & $95.0_{\pm 6.4}$  & $99.9_{\pm 0.1}$  & $130.3_{\pm 2.3}$  & $1872.5_{\pm 50.3}$ & 76 & 259 \\
\midrule
R-GCN$^c$                & AIFB     & $91.7_{\pm 0.0}$  & $95.0_{\pm 3.0}$  & $595.4_{\pm 15.8}$ & $18.7_{\pm 0.6}$   & 20 & 219 \\
\bottomrule
\end{tabular}
\\[0.4em]
\raggedright\scriptsize
$^a$RelaNN and the baseline both use the Paper$\to$Author scope of HGT~\cite{hu2020heterogeneous} (one edge type per layer). Full-scope PyG and pyHGT baselines, traversing all six edge types, run $1.5$--$4\times$ slower.
$^b$The \texttt{gear/dhn} baseline runs at the configuration of DHN~\cite{maehara2024deep}, reproducing its reported accuracy.
$^c$\texttt{torch-rgcn} and RelaNN both use full per-relation R-GCN weights (24M parameters, no basis decomposition), so the comparison is parameter-matched.
\end{table}

%\paragraph{Matching reference accuracy.}
On GCN and HGT, RelaNN reaches accuracy identical to the baseline, confirming that the compiled execution is faithful to the reference.
The other comparisons run from independent initialization: HyGNN lands within one standard deviation of the baseline on both datasets, and R-GCN reaches $91.7\%$ on AIFB, a few points below the parameter-matched \texttt{torch-rgcn} reference.
For DHN~\cite{maehara2024deep} we follow its two pattern sets and reproduce its results at both: the smaller set (C2:4) is insufficient for the task, while with the larger set (C2:10) RelaNN reaches $100\%$ on CSL and $95\%$ on EXP.

\paragraph{HGT: attention as rules.}
Figure~\ref{fig:hgt-attention-correspondence} places HGT's attention head beside its RelaNN encoding: each of the equations defining $K^i$ and $Q^i$ become a single rule, and the $\mathrm{ATT}^i$ equation becomes two (a dot-product rule and a softmax rule).
%, with matching variable names and subscripts.
The same correspondence holds across the architectures we implemented: each source equation maps to just one or two rules.
%, so a RelaNN program can be checked against its paper line by line.
%It also keeps architectural edits local: changing an aggregation function or adding an embedded relation is a single-rule edit, whereas the equivalent change in a hand-written PyG implementation typically touches module boundaries, parameter dictionaries, and message-passing hooks.
% A full equation-to-rule listing for HGT and DHN appears in Appendix Figures~\ref{fig:hgt-latex-vs-RelaNN} and~\ref{fig:dhn-latex-vs-RelaNN}.
% \yl{A full equation-to-rule listing for HGT and DHN is included in the supplementary material.}

% \footnote{\url{https://github.com/yuvallu/relann}}

\begin{figure}[b]
\vskip-1em
\centering
\hrule
\begin{minipage}[t]{\linewidth}
{\small
\begin{gather*}
K^i(s) = \mathbf{KLin}^i_{\tau(s)}\!\bigl(H^{(L-1)}[s]\bigr), \quad Q^i(t) = \mathbf{QLin}^i_{\tau(t)}\!\bigl(H^{(L-1)}[t]\bigr) \\
\mathrm{ATT}^i = \mathrm{softmax}_t\!\left(K^i W^{\mathrm{ATT}}_{\phi(e)} {Q^i}^\top \!\cdot\! \mu \,/\, \sqrt{d/h}\right)
\end{gather*}
}
\hrule
%\smallskip
\begin{code}
\rn{K}(\ca{s}; \fn{KLin}\tp{<L,tau_s,i>}(\ea{z}))  \neck \rn{HGT}\tp{<tau_s,L-1>}(\ca{s}; \ea{z}) .
\rn{Q}(\ca{t}; \fn{QLin}\tp{<L,tau_t,i>}(\ea{z}))  \neck \rn{HGT}\tp{<tau_t,L-1>}(\ca{t}; \ea{z}) .
\rn{Dot}(\ca{s},\ca{t}; (\ea{z_k} @ \fn{W_ATT}\tp{<L,phi,i>} @ \ea{z_q}\fn{.T}) * \fn{Mu}/\fn{sqrt}(\nm{d}/\nm{h})) \neck
	\rn{K}(\ca{s}; \ea{z_k}), \rn{phi}(\ca{s},\ca{t}), \rn{Q}(\ca{t}; \ea{z_q}) .
\rn{ATT}(\ca{s},\ca{t}; \ea{z}) \neck \fn{Softmax}(\rn{Dot})(\ca{s},\ca{t}; \ea{z}) .
\end{code}
\hrule
\end{minipage}
\vskip-1em
\caption{Equation-to-rule correspondence for the HGT attention head~\cite{hu2020heterogeneous}. Each equation (top) maps to one or two RelaNN rules.
%, preserving variable names and subscripts. 
$\mathrm{KLin}$ and $\mathrm{QLin}$ are the key and query layers.}
\label{fig:hgt-attention-correspondence}
\vskip-1em
\end{figure}

\paragraph{DHN: cycles as joins.}
%\ylinline{Consider to make shorter, even remove the code example.}
DHN's core construct is graph homomorphism enumeration, which detects subgraph patterns and relates to the $k$-WL hierarchy~\cite{maron2019provably}.
This enumeration has no native layer in PyG or DGL~\cite{wang2019dgl}, so published implementations embed it in several hundred lines of custom tensor code.
RelaNN expresses it declaratively in a few rules: a triangle homomorphism becomes a cyclic self-join over edge-like relations.
\begin{code}
\rn{C3_Agg}(\ca{g},\ca{n}; \fn{sum}(
  \fn{Mu}\tp{<'C3',0>}(\ea{h_n}) * \fn{Mu}\tp{<'C3',1>}(\ea{h_v}) * \fn{Mu}\tp{<'C3',2>}(\ea{h_w})) ) \neck
    \rn{Edge}(\ca{g},\ca{n},\ca{v}), \rn{Edge}(\ca{g},\ca{v},\ca{w}), \rn{Edge}(\ca{g},\ca{w},\ca{n}),
    \rn{H0}(\ca{g},\ca{n}; \ea{h_n}), \rn{H0}(\ca{g},\ca{v}; \ea{h_v}), \rn{H0}(\ca{g},\ca{w}; \ea{h_w}) .
\end{code}
The three \rn{Edge} body relations form the triangle, the \rn{H0} relations fetch each matched node's previous-layer embedding, and the \fn{Mu} transformations apply a per-position learned MLP defined with templates (Section~\ref{sec:lang-extensions}).
The \fn{sum} aggregates over all triangle matches anchored at node \ca{n}.
This single rule is a direct transcription of the per-pattern term in DHN~\cite{maehara2024deep}.
% \yl{TODO: fill in the REPO-URL in the footnote below with the real repository link.}
% \yl{The full programs (21 lines for C2:4, 76 lines for the templated C2:10 variant) are provided alongside the reference code as supplementary material.\footnote{\url{https://github.com/REPO-URL}}}
%The full programs (21 lines for C2:4, 76 lines for the templated C2:10 variant) are provided alongside the reference code in the project repository.
% \footnote{\url{https://github.com/yuvallu/relann}}
A full equation-to-rule listing for HGT and DHN is available in the project repository.

\paragraph{HyGNN: hyperedges as relations.}
HyGNN predicts drug-drug interactions by modeling each drug as a hyperedge over its molecular substructures: the substructures are the nodes of a hypergraph, and each drug is a hyperedge grouping the substructures it is composed of.
Standard GNN APIs represent an edge as a pair of nodes and have no first-class encoding for the many-to-one incidence between substructures and the drug they belong to.
PyG does ship a generic \texttt{HypergraphConv}, but it implements a different message-passing scheme and is not equivalent to HyGNN's double-attention mechanism.\footnote{\url{https://pytorch-geometric.readthedocs.io/en/2.5.0/generated/torch_geometric.nn.conv.HypergraphConv.html}}
In RelaNN, the set of (drug, substructure) incidences is an ordinary two-column relation, and message passing between a drug-hyperedge and its substructure-nodes is a natural join on the shared substructure identifier, with no special hypergraph API.

%% ===================================================================
\subsection{Runtime}
\label{sec:exp-runtime}
%% ===================================================================

Table~\ref{tab:unified-results} reports per-epoch training time.
On ten of the eleven rows, the per-epoch time is close to the baseline or well below it: RelaNN stays within roughly $2\times$ on GCN and HGT, and on the four DHN rows, it is $2$--$14\times$ faster.
These DHN margins partly reflect that \texttt{gear/dhn} is a research prototype rather than a tuned library such as PyG.
The single pronounced outlier is R-GCN on AIFB, where RelaNN is roughly $30\times$ slower than the specialized \texttt{torch-rgcn} baseline.
RelaNN's current implementation is an unoptimized proof of concept, and that row is where the gap is widest.
Even so, these numbers already show that a fully declarative, database-style pipeline trains published architectures at a practical per-epoch cost.
We leave cost-based query optimization to future work (Section~\ref{sec:discussion}).
% ref retargeted from the removed sec:future-queryplan label
%\dl{This section is begging to get asked "so why didnt you implement this" all you should say here is that we are ranging between 0.aX-bX compared to the baselines, showing the promise of our apporach. State that this is a poc system and point to future work. If there is some interesting insights on the big variance, you can explain it shortly, but dont assert that this is an artifact of some part of the system and not another. This assertion is begging to be asked to be supported with an ablation.}

%% ===================================================================
\subsection{Case Study: Task-Specific Architecture}
\label{sec:exp-design}
%% ===================================================================

The previous studies implement existing graph architectures.
This case study illustrates the other direction: a data owner starts from a relational database, designs a task-specific architecture in RelaNN, and refines it through local rule-level edits.

\subsubsection*{Task.}
We use four tables from the rel-f1 database~\cite{fey2024relational} which are extended into embedded relations: \rel{Drivers}$/1\langle$\nm{d\_dr}$\rangle$, each driver belonging to a team in \rel{Constructors}$/1\langle$\nm{d\_co}$\rangle$, with each team taking part in \rel{Races}$/1\langle$\nm{d\_ra}$\rangle$, with one row per driver per race in \rel{Results}$/4\langle$\nm{d\_re}+\nm{h}*\nm{2}$\rangle$, where \nm{d\_dr}, \nm{d\_co}, \nm{d\_ra}, \nm{h} are predefined using aliases.
The rules operate on these tables directly---there is no graph to construct.
The binary \texttt{driver-dnf} task predicts whether a driver receives a Did-Not-Finish in the next race period, evaluated by the area under the ROC curve (AUROC).

\paragraph{Initial architecture.}
We start from an architecture that embeds each \rel{Results} row together with its team and race context, averages these embeddings per driver into a history, and scores the driver from its own embedding and that history.
The full six-rule program appears below.
It reaches 0.610 AUROC, above the 0.5 of a random classifier but below the RelBench baselines for this task: 0.686 for a LightGBM model over the entity table and 0.726 for a GNN~\cite{fey2024relational}.
\begin{listing}[htbp]
\begin{code}[numbers=left]
\rn{DriverEmb}(\ca{did}; \fn{ReLU}(\fn{Linear}(\nm{d_dr}, \nm{h})(\ea{z}))) \neck \rn{Drivers}(\ca{did}; \ea{z}).
\rn{ConsEmb}(\ca{cid}; \fn{ReLU}(\fn{Linear}(\nm{d_co}, \nm{h})(\ea{z}))) \neck
    \rn{Constructors}(\ca{cid}; \ea{z}) .
\rn{RaceEmb}(\ca{rid}; \fn{ReLU}(\fn{Linear}(\nm{d_ra}, \nm{h})(\ea{z}))) \neck \rn{Races}(\ca{rid}; \ea{z}) .
\fn{ResultProj} = \fn{ReLU}(\fn{Linear}(\nm{d_re}+\nm{h}*\nm{2}, \nm{h})) .
\rn{ResultEmb}(\ca{eid},\ca{did}; \fn{ResultProj}(\fn{Concat}(\ea{z_r}, \ea{z_c}, \ea{z_ra}))) \neck
    \rn{Results}(\ca{eid},\ca{did},\ca{rid},\ca{cid}; \ea{z_r}), \rn{ConsEmb}(\ca{cid}; \ea{z_c}),
    \rn{RaceEmb}(\ca{rid}; \ea{z_ra}) .
\rn{History}(\ca{did}; \fn{mean}(\ea{z_e})) \neck \rn{ResultEmb}(\ca{eid},\ca{did}; \ea{z_e}) .
\rn{Score}(\ca{did}; \fn{Linear}(\nm{h}*\nm{2}, \nm{1})(\fn{Concat}(\ea{z_d}, \ea{z_h}))) \neck
    \rn{DriverEmb}(\ca{did}; \ea{z_d}), \rn{History}(\ca{did}; \ea{z_h}) .
\end{code}
%\caption{The initial six-rule RelaNN program for the \texttt{driver-dnf} task: four embedding rules, one history aggregation, and a final score.
%\texttt{d\_dr}, \texttt{d\_co}, \texttt{d\_ra}, \texttt{d\_re} are the input feature dimensions, and \texttt{h}~=~16 is the hidden dimension.}
% \label{code:driver-task-specific}
\end{listing}
\paragraph{Refining the architecture.}
The initial \rn{History} rule averages a driver's past results with a uniform \fn{mean}, weighting every result equally.
A natural refinement is to let the model decide how much each result counts, with a learned $[0,1]$ gate per result:
\begin{code}
\fn{WAvg} = \fn{Sigmoid}(\fn{Linear}(\nm{d_re}+\nm{h}*\nm{2}, \nm{1})) .
\rn{ResultGate}(\ca{eid},\ca{did}; \fn{WAvg}(\fn{Concat}(\ea{z_r}, \ea{z_c}, \ea{z_ra}))) \neck
    \rn{Results}(\ca{eid},\ca{did},\ca{rid},\ca{cid}; \ea{z_r}), \rn{ConsEmb}(\ca{cid}; \ea{z_c}),
    \rn{RaceEmb}(\ca{rid}; \ea{z_ra}) .
\end{code}
We add this \rn{ResultGate} rule and replace the \rn{History} rule so that it averages the gated embeddings:
\begin{code}
\rn{History}(\ca{did}; \fn{mean}(\ea{z_e} * \ea{z_w})) \neck \rn{ResultEmb}(\ca{eid},\ca{did}; \ea{z_e}),\\
    \rn{ResultGate}(\ca{eid},\ca{did}; \ea{z_w}) .
\end{code}
This one new rule and one edited rule lift performance to 0.707 AUROC, past the LightGBM baseline and close to the GNN baseline.
As aggregation is a first-class rule-level construct, the edit requires no change to index bookkeeping, parameter allocation, or gradient plumbing: the compiler re-derives all of them from the edited rules.

% % Full-width figure: two-column body is too narrow for this listing
% \begin{figure*}[t]
% \centering
% \begin{code}
% \rn{DriverEmb}(\ca{did}; \fn{ReLU}(\fn{Linear}(\nm{d_dr}, \nm{h})(\ea{z}))) \neck \rn{Drivers}(\ca{did}; \ea{z}) .
% \rn{ConsEmb}(\ca{cid}; \fn{ReLU}(\fn{Linear}(\nm{d_co}, \nm{h})(\ea{z}))) \neck \rn{Constructors}(\ca{cid}; \ea{z}) .
% \rn{RaceEmb}(\ca{rid}; \fn{ReLU}(\fn{Linear}(\nm{d_ra}, \nm{h})(\ea{z}))) \neck \rn{Races}(\ca{rid}; \ea{z}) .
% \rn{ResultEmb}(\ca{eid},\ca{did}; \fn{ReLU}(\fn{Linear}(\nm{d_re}+\nm{h}*\nm{2}, \nm{h})(\fn{Concat}(\ea{z_r}, \ea{z_c}, \ea{z_ra})))) \neck
%         \rn{Results}(\ca{eid},\ca{did},\ca{rid},\ca{cid}; \ea{z_r}), \rn{ConsEmb}(\ca{cid}; \ea{z_c}), \rn{RaceEmb}(\ca{rid}; \ea{z_ra}) .
% \rn{History}(\ca{did}; \fn{mean}(\ea{z_e})) \neck \rn{ResultEmb}(\ca{eid},\ca{did}; \ea{z_e}) .
% \rn{Score}(\ca{did}; \fn{Linear}(\nm{h}*\nm{2}, \nm{1})(\fn{Concat}(\ea{z_d}, \ea{z_h}))) \neck \rn{DriverEmb}(\ca{did}; \ea{z_d}), \rn{History}(\ca{did}; \ea{z_h}) .
% \end{code}
% \caption{The initial six-rule RelaNN program for the \texttt{driver-dnf} task: four embedding rules, one history aggregation, and a final score.
% \texttt{d\_dr}, \texttt{d\_co}, \texttt{d\_ra}, \texttt{d\_re} are the input feature dimensions, and \texttt{h}~=~16 is the hidden dimension.}
% \label{fig:f1-gated-history}
% \end{figure*}

\section{Conclusions and Future Work}
\label{sec:discussion}
% In this paper, we \dl{revisited? who visited this first? try to stay close to the messaging in the abstract and the results.} \sout{revisited the integration of databases and deep learning from a declarative perspective}\yl{proposed a declarative approach to deep learning over relational data}, arguing that the prevailing graph-based, imperative approach introduces unnecessary complexity by separating data management from model design. Our proposed RelaNN enables the design of neural architectures directly over relational data through an extension of the relational model, formalized as NRA. By lifting relational operators to manipulate tuple embeddings and coupling them with embedding aggregation, transformations, and training, RelaNN unifies query processing, model specification, and optimization within a single framework. Our empirical study demonstrates that this abstraction is not only \e{expressive}---capturing a wide range of modern graph-learning architectures---but also \e{practical}, achieving \sout{competitive accuracy and runtime}\yl{comparable accuracy and competitive runtime} while significantly simplifying development. Hence, it illustrates that declarative query languages can serve as a natural and powerful foundation for deep learning over relational data, opening new directions for integrating learning and data management systems. \dl{ This is a bit too long, especially the "by lifting...." try to say concisly were we started from, what we accomplished, how we did it and what is the implication.}
Deep learning over relational databases conventionally requires interfacing with external systems beyond the database, incurring non-trivial overhead. We introduced NRA, a new algebra that extends the relational model with learned embeddings. NRA naturally merges relational and tensor algebra, enabling deep learning models to be defined directly and elegantly over the database. We presented %\sa{a proof of concept in the form of}
RelaNN, a declarative system and language over NRA. We have shown that RelaNN enables the implementation of %allows us to implement 
a wide variety of SOTA neural architectures, in a fraction of the complexity and lines of code of the original implementations. While our implementation lags behind enterprise systems in terms of training runtime, it demonstrates a major step
%way forward 
towards making learning over relational data as simple as writing queries.

To make RelaNN useful for real-world scenarios, future work will focus on two main areas.
%1) scaling to out of memory databases and 2) query optimization. 
The first is \textit{scaling for memory}.
%\textit{scaling out of GPU memory:}}
In this paper we focused on datasets that fit in the memory of modern GPUs. Real-world datasets often exceed this limit. Handling such datasets requires database sampling~\cite{10.5555/3294771.3294869,DBLP:conf/kdd/ChiangLSLBH19,Zeng2019GraphSAINTGS} and streaming the data in mini-batches from the database to the GPU~\cite{khan2025sqlgnn} to optimize performance and maintain accuracy.
The second is \textit{query optimization}. RelaNN, being a DBMS that separates specification from logical and physical planning, is amenable to extensive optimizations using term-rewriting techniques. It will be interesting to explore three types of optimizations: %In future work we will explore three types of optimizations: 
1) Relational optimizations, such as selection pushdowns and join reordering, that can be delegated
%pushed \sa{a different word? to not repeat push again.} 
from the NRA operators to the database via SQL. 2) Tensor optimizations that can fuse consecutive embedding operators into a single GPU kernel. 3) Novel hybrid optimizations that take advantage of the interplay between the content and the embedding. For example, it would be interesting to apply ideas~\cite{kumar2015glm,chen2017morpheus,olteanu2020borg} involving pushing down transformations before joins so that they operate on embeddings once per node, instead of once per edge.

% The recent emergence of \e{relational foundation models}~\cite{DBLP:journals/tmlr/FranksECWSFK25,DBLP:conf/icml/WangWGWYWZ25,DBLP:journals/corr/abs-2604-12596} holds the promise of learning from a vast history of many databases, with varying levels of relevance to the user's current database and task.
% %at hand,
% Such models have been shown to accelerate learning and boost accuracy.
% % thereby considerably accelerating learning and boosting accuracy.
The recent emergence of \e{relational foundation models}~\cite{DBLP:journals/tmlr/FranksECWSFK25,DBLP:conf/icml/WangWGWYWZ25,DBLP:journals/corr/abs-2604-12596}, pretrained across many databases, promises to accelerate learning and improve accuracy.
We plan to investigate the implementation of foundation models in RelaNN. 
\bibliographystyle{ACM-Reference-Format}
\bibliography{main}

\end{document}